\documentclass[12pt, letterpaper]{article} 
\usepackage[utf8]{inputenc}
\usepackage[paperwidth=216mm, paperheight=279mm, margin=1.87cm, marginparsep=0pt]{geometry}
\usepackage{courier} 
\usepackage{hyperref}
\usepackage[round]{natbib}
\setcitestyle{authoryear,open={(},close={)}}
\usepackage{graphicx}
\usepackage{placeins}
 \usepackage{multirow}
 \usepackage{amsmath}
 \usepackage{float}
 \usepackage{amsfonts}
 \usepackage[affil-it,blocks]{authblk}
 \usepackage{chngcntr}
 \usepackage{subcaption}
\usepackage{algorithm}
\usepackage[noend]{algpseudocode}
\allowdisplaybreaks
\usepackage{attachfile2}
\usepackage{xurl}
\usepackage{hyperref} 
\usepackage[defaultcolor=black]{changes}
\usepackage[
singlelinecheck=false 
]{caption}
\usepackage{xr}
\usepackage{authblk}

\providecommand{\keywords}[1]
{
  \small	
  \noindent \textbf{Keywords:} #1
}

\title{A switching state-space transmission model for tracking epidemics and assessing interventions}
\author[1]{Jingxue Feng \thanks{Email address: jingxuef@sfu.ca}}
\author[1]{Liangliang Wang \thanks{Corresponding author. Email address: lwa68@sfu.ca; Postal address: 8888 University Dr W, Burnaby, BC V5A 1S6 }}
\affil[1]{Department of Statistics and Actuarial Science, Simon Fraser University, Burnaby, BC, Canada V5A1S6}
\date{}

\begin{document}
\maketitle

\begin{abstract}
    The effective control of infectious diseases relies on accurate assessment of the impact of interventions, which is often hindered by the complex dynamics of the spread of disease. A Beta-Dirichlet switching state-space transmission model is proposed to track underlying dynamics of disease and evaluate the effectiveness of interventions simultaneously. As time evolves, the switching mechanism introduced in the susceptible-exposed-infected-recovered (SEIR) model is able to capture the timing and magnitude of changes in the transmission rate due to the effectiveness of control measures. The implementation of this model is based on a particle Markov Chain Monte Carlo algorithm, which can estimate the time evolution of SEIR states, switching states, and high-dimensional parameters efficiently. The efficacy of the proposed model and estimation procedure are demonstrated through simulation studies. With a real-world application to British Columbia's COVID-19 outbreak, the proposed switching state-space transmission model quantifies the reduction of transmission rate following interventions.  The proposed model provides a promising tool to inform public health policies aimed at studying the underlying dynamics and evaluating the effectiveness of interventions during the spread of the disease.
\end{abstract}

\begin{quotation}
\end{quotation}

\newpage
\section{Introduction}
\label{sec: Introduction}
        
 The Coronavirus disease 2019 (COVID-19) pandemic, caused by severe acute respiratory syndrome coronavirus 2 (SARS-CoV-2), was first reported in China in December 2019 and lasted for over three years globally. As of May 2023, the pandemic has resulted in more than 0.7 billion confirmed cases and 6.9 million fatalities \citep{who-covid19-dashboard}. Various types of interventions have been introduced throughout the course of pandemic to prevent or mitigate the spread of the disease at regional levels, such as social distancing, closure of public places, and implementation of vaccines. Understanding the effectiveness of these interventions during the course of pandemic is crucial in managing the spread of future diseases. Although several dynamic epidemiological models have been proposed for tracking epidemics and assessing interventions  \citep{dukic2012tracking, osthus2017forecasting, asher2018forecasting, dehning2020inferring, wang2020epidemiological}, the challenge lies in developing a more flexible and generalized method that can simultaneously track the underlying dynamics of the disease and evaluate the effectiveness of interventions.

\cite{dukic2012tracking}, \added{\cite{asher2018forecasting}} and \cite{osthus2017forecasting} combined compartmental epidemiological models with a state-space framework for estimating disease dynamics. However, they cannot provide insights closely tied to external intervention effects, such as public health measures, during the process of disease dynamics. \cite{wang2020epidemiological} and \cite{dehning2020inferring} extended similar state-space models to detect change points due to publicly announced interventions in the spread of COVID-19. \cite{wang2020epidemiological} extended the state-space susceptible-infectious-recovered (SIR) model \citep{osthus2017forecasting} by incorporating time-varying quarantine protocols. They modified the transmission rate based on specific quarantine protocols implemented in a region, but their transmission rate modifier was not informed by data. Instead, it was treated as a fixed quantity, either a step function with pre-specified change time points or a continuous exponential function with a certain rate. To provide accurate measurements, uncertainty about the transmission rate modifier needs to be accounted for. \cite{dehning2020inferring} used a state-space SIR model to infer change points caused by governmental interventions. However, their model is limited by assuming that each mitigation measure leads to approximately a $ 50\%$ reduction in the transmission rate, and it is also constrained by quite informative priors for the time of change points.

We propose a Beta-Dirichlet switching state-space model combined with the susceptible-infectious-exposed-recovered dynamics (BDSSS-SEIR) model to track the underlying disease dynamics and detect changes of regimes. The proposed model is motivated by \cite{osthus2017forecasting}, in which a Beta-Dirichlet state-space SIR model is solely used to forecast seasonal influenza. We introduce a regime switching mechanism to assess the impacts of external interventions by dynamically capturing the timing and magnitude of changes in the transmission rate. Instead of fitting the model using the Monte Carlo Markov Chain (MCMC) algorithm, as was done in \cite{osthus2017forecasting}, we employ a more advanced Bayesian inference method called particle MCMC \citep{andrieu2010particle}. This method allows us to efficiently explore the latent trajectories and high-dimensional parameter space in our proposed model. Particle MCMC constructs efficient high-dimensional proposal distributions via Sequential Monce Carlo (SMC) in the MCMC iterations. Therefore, it reduces computational difficulties in integrating out the time-evolving latent variables while exploring the target distribution. We adapt one of the particle MCMC methods, specifically particle Gibbs sampling \citep{andrieu2010particle}, to perform inference in our proposed model.  
For all sampling processes in simulation studies and real data analysis, 
we utilized the \texttt{R} software program, version 3.6.3 \citep{R}. 
All source codes are publicly available at
\url{https://github.com/SFU-Stat-ML/BDSSS-SEIR}.

The advantages of our proposed model are summarized as follows. First, the proposed model uses a regime switching mechanism to dynamically capture the timing and magnitude of intervention effects. This approach is more compact and efficient than either relying on informative priors on the time of change points \citep{dehning2020inferring} or using a fixed quantity to represent changes in transmission rate \citep{wang2020epidemiological}. Second, our model incorporates uncertainty into the reduction of transmission rate across different regimes, providing more accurate results compared to making assumptions on the reduction beforehand \citep{dehning2020inferring}. Third, our model allows the transmission rate to rise and fall as policies change, enabling more flexibility while assessing interventions. To the best of our knowledge, this is the first study to incorporate a switching state-space model with a compartmental model in studying disease dynamics. We believe that our approach will be valuable for better understanding and combating infectious diseases.

The rest of the paper is organised as follows. In Section \ref{section: Methodology}, we introduce the Beta-Dirichlet switching state-space SEIR model and describe the particle MCMC algorithm used in our proposed model. Section \ref{section: Results} presents the results of simulation studies conducted under two-regime and three-regime scenarios, respectively. Furthermore, we demonstrate the proposed BDSSS-SEIR model on weekly active case count data in British Columbia, Canada. Finally, in Section \ref{section: Discussion}, we conclude the study and discuss the potential challenges associated with our model.

\section{Methodology} 
\label{section: Methodology}
Our proposed BDSSS-SEIR model embeds the SEIR model in a switching state-space framework to capture the effectiveness of interventions during epidemics over time. To better understand the model dynamics, we first introduce a modified SEIR system governed by a switching state variable in the dynamic model. Then, we illustrate the probabilistic structure of the BDSSS-SEIR model. Finally, we provide detailed information on the particle MCMC algorithm, which combines the strengths of SMC and MCMC to estimate model parameters and latent variables.

\subsection{Modified susceptible-exposed-infected-recovered (SEIR) model} 
\label{sec: Modified SEIR Model} 
We first present a classic SEIR model for a directly transmitted infectious disease before introducing the modified SEIR model, which is a key component of the probabilistic switching state-space model. The classic SEIR model assumes that individuals transition between four compartments --- susceptible (S), exposed (E), infected (I), and recovered (R) --- over time \citep{anderson1991infectious}. We assume a constant total population size of 1 over time, \added{allowing us to focus on the relative proportions of individuals in different compartments.} The classic SEIR model in epidemiology, describing the spread of infectious disease, is represented by the following set of nonlinear ordinary differential equations:
\begin{equation}
\label{eq:SEIRsystem}
    \begin{aligned}
    \frac{dS}{dt} = - \beta S_t I_t, \  \
    \frac{dE}{dt} = \beta S_t I_t - \alpha E_t, \  \
    \frac{dI}{dt} = \alpha E_t - \gamma I_t, \ \
   \text{and }  \frac{dR}{dt} = \gamma I_t,\
    \end{aligned}
\end{equation}
 where [$\frac{dS}{dt}, \frac{dE}{dt}, \frac{dI}{dt}, \frac{dR}{dt}$] are the time derivatives of the susceptible, exposed, infected, and recovered compartments at time $t$, and [$S_t, E_t, I_t, R_t$] represents the number of individuals in each compartment at time $t$. In this model, $\beta$ is the disease transmission rate, $\alpha$ is the latency rate at which exposed individuals become infected but not yet infectious, and $\gamma$ is the recovery rate that at which infected individuals recover and become immune to the disease. Specifically,  individuals in the susceptible compartment become exposed at a rate of $\beta S_t I_t$. The exposed individuals do not yet show symptoms, but they become infectious after a latent period $1/\alpha$ (i.e., the time between infection and the onset of infectiousness). Infectious individuals remain infectious for a period $1/\gamma$ before recovery. The classic SEIR model in Equation (\ref{eq:SEIRsystem}) assumes a constant population size, with $S_t+E_t+I_t+R_t=1$ for all $t$, indicating that birth and death rates are equal.

To capture the dynamic impact of intervention measures on the transmission dynamics of the disease, we introduce a time-varying parameter $f_{x_t} \in (0,1]$ into the classic SEIR model,  which represents a transmission rate modifier associated with a switching regime $x_t \in \{1,2,\ldots,K\}$ at time $t$. The switching state variable $x_t$ is characterized by its initial distribution $\mu_\psi(x_1)$ and evolves according to a $K$-state first-order Markov switching process with transition probabilities 
\begin{equation}
\label{eq:transitionProb}
    P(x_t=j|x_{t-1}=i) = \pi_{ij}, \ \forall i,j \in \{1,\ldots,K \},
\end{equation} 
where $\sum_{j=1}^K \pi_{ij} = 1$. \added{This structure of transition probabilities was first introduced by \cite{hamilton1989new} and has been combined with state-space models in various studies \citep{kim1994dynamic, whiteley2010efficient, kim2017state}.} The value of $x_t$ determines how the transmission rate is modified at time $t$. For example, if $K=2$, we may define transmission rate modifier as
$$f_{x_t}=
    \begin{cases}
      1, & \text{if}\ x_t=1, \\
      0.5, & \text{if}\ x_t = 2.
    \end{cases}$$
In this case, the transmissibility of virus remains at its original baseline in the first regime, while it is reduced by 50\,\% due to external interventions in the second regime. By incorporating $x_t$ into $f_{x_t}$, we account for the time-varying nature of the effectiveness of intervention measures. Estimating $x_t$ is not only important to estimate the regime of epidemics, but also crucial for understanding the effects of interventions at time $t$. In our proposed BDSSS-SEIR model, we assume that the transmission rate is always kept at the baseline level in the first regime. In other regimes, the external interventions impact the transmission rate to various extents. Uncertainties are therefore added to $f_{x_t}$ when $x_t \ne 1$, \added{as illustrated in Section \ref{sec: prior on parameters}}. This switching state variable $x_t$ governs a modified SEIR system as follows
\begin{equation}
\label{eq:modifiedSEIRsystem}
    \begin{aligned}
    \frac{dS}{dt} = - f_{x_t} \beta S_t I_t, \ \
    \frac{dE}{dt} = f_{x_t} \beta S_t I_t - \alpha E_t,  \ \
    \frac{dI}{dt} = \alpha E_t - \gamma I_t, \ \
   \text{ and }
    \frac{dR}{dt} =  \gamma I_t,
    \end{aligned}
\end{equation}
where the transmission rate $\beta$ is adjusted by $f_{x_t}$ as time progresses. This transmission rate modifier can incorporate the influence of various factors and interventions implemented over time, such as social distancing measures, mask-wearing policies, or vaccination campaigns.

\begin{figure}[ht]
    \centering
    \begin{subfigure}[b]{0.49\textwidth}
        \includegraphics[width=\textwidth]{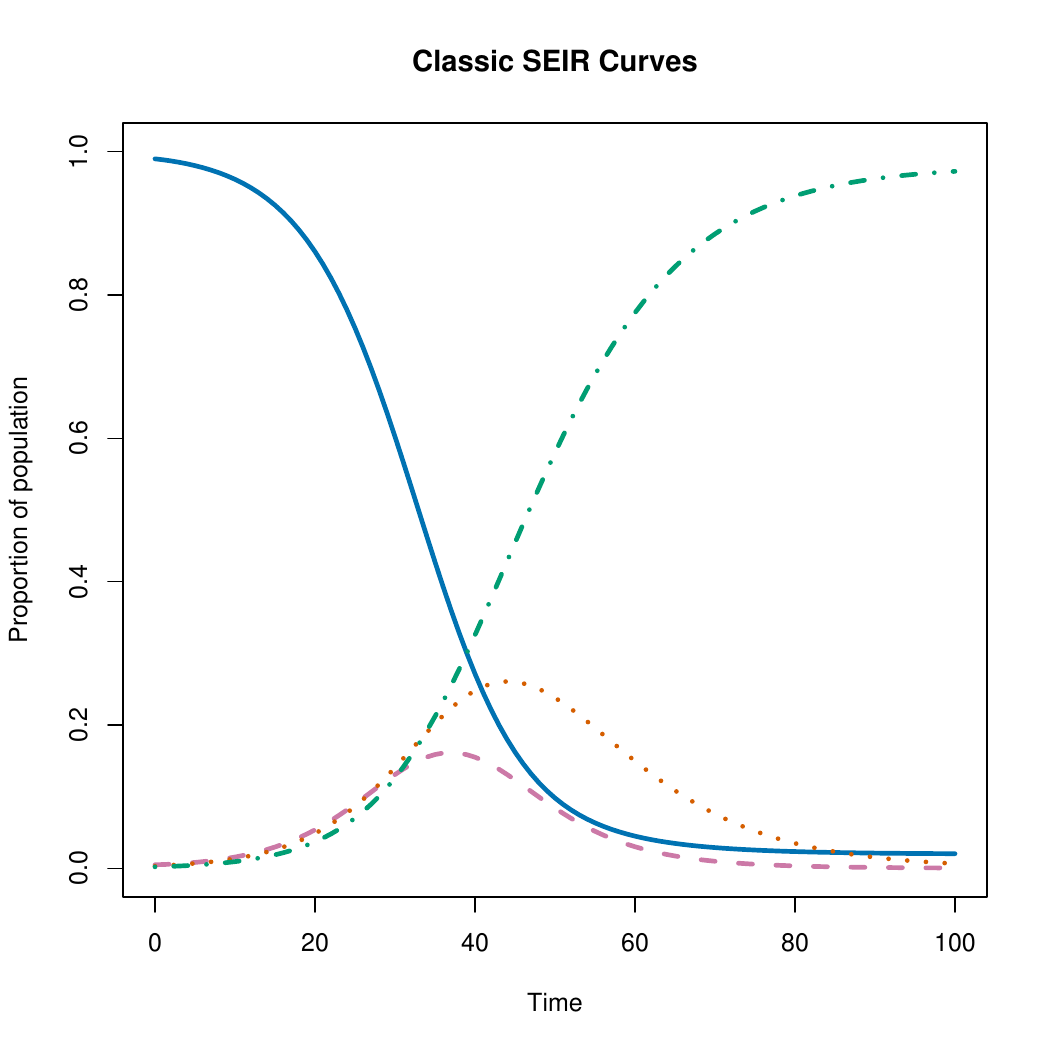}
        \caption{}
        \label{fig:SEIR Curves}
    \end{subfigure}
    \hfill
    \begin{subfigure}[b]{0.49\textwidth}
        \includegraphics[width=\textwidth]{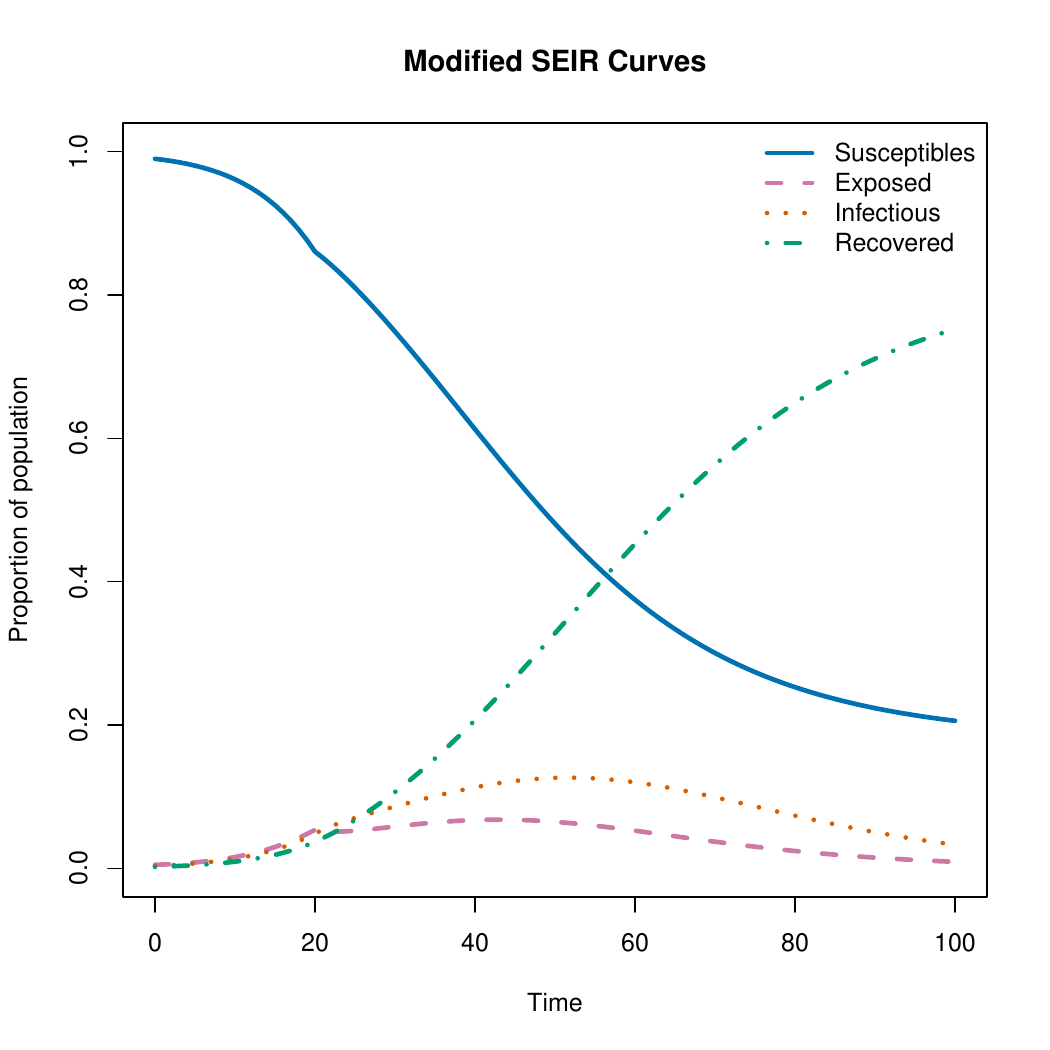}
        \caption{}
        \label{fig:SEIR Curves with Transmission Modifier}
    \end{subfigure}
    \caption{(a) Simulated SEIR curves with $S_1$ = 0.99, $E_1$ = 0.005, $I_1$ = 0.003, $R_1$ = 0.002, $\beta$ = 0.4, $\alpha$ = 0.2, and $\gamma$ = 0.1. (b) Modified SEIR curves with $S_1$ = 0.99, $E_1$ = 0.005, $I_1$ = 0.003, $R_1$ = 0.002, $\beta$ = 0.4, $\alpha$ = 0.2, $\gamma$ = 0.1, plus a transmission rate modifier $f_{x_t}$ such that $f_{x_t} = 1$ for $t=1,\ldots,20$, and $f_{x_t} = 0.5$ for $t=21,\ldots,100$. }
    \label{fig:both_figures}
\end{figure}

For the purpose of comparison, we simulate classic SEIR curves and modified SEIR curves for 100 time points, respectively, in Figure \ref{fig:both_figures}. All parameter values and initial values of SEIR dynamics are the same in both scenarios. However, in Figure \ref{fig:SEIR Curves with Transmission Modifier}, the transmission rate is kept at the baseline level for $t=1,\ldots,20$, and then reduced by 50\,\% for $t=21,\ldots,100$ due to intervention measures. It can be observed that the number of infected cases rises rapidly and reaches a peak within a short period if no preventive measures are taken. However, if preventive measures are implemented, resulting in a reduction in the transmission rate, the infectious curve is flattened. At the end of the epidemic cycle in the modified SEIR system, approximately 20\,\% of the population remains unexposed to the disease. By incorporating the time-varying parameter $f_{x_t}$ into the classic SEIR model, the modified SEIR system demonstrates the benefits of implementing interventions to reduce the transmission rate and flatten the epidemic curve.

\subsection{Beta-Dirichlet switching state-space SEIR model}
\label{sec: BDSSS-SEIR Model}

We introduce the probabilistic structure of the switching state-space model motivated by the modified SEIR system in Section \ref{sec: Modified SEIR Model}. The dynamics of the disease are described by the evolution of the latent SEIR states, $\boldsymbol{\theta}_t = [S_t, E_t, I_t, R_t]^\top$, which always sums to unity. Throughout this paper, we use a bold symbol to represent a vector or matrix. In the transition equation, the four-dimensional Markov process $\boldsymbol{\theta}_t$ is assumed to follow
\begin{equation}
    \begin{aligned}
    \label{eq: BDSSS-SEIR-Transition Equation}
    \boldsymbol{\theta}_t &| \boldsymbol{\theta}_{t-1},x_t \text{\added{, $\psi$}} \sim \text{Dirichlet}(\kappa  r(\boldsymbol{\theta}_{t-1}; \alpha, \beta, \gamma, f_{x_t})),
    \end{aligned}
\end{equation}
where the latent state $\boldsymbol{\theta}_1$ comes from an initial distribution $\mu_\psi(\boldsymbol{\theta}_1|x_1)$, and the underlying proportions of the susceptible, exposed, infectious, and recovered population at time $t>1$ depend on the previous state $\boldsymbol{\theta}_{t-1}$, the current regime of the model $x_t$,  \added{and the model parameters $\psi$}.  The parameter $\kappa$ controls the variability of $\boldsymbol{\theta}_t$ over time. The function $r(\boldsymbol{\theta}_{t-1}; \alpha, \beta, \gamma, f_{x_t})$ represents a discretized solution to the modified SEIR system in Equation (\ref{eq:modifiedSEIRsystem}), starting the ordinary differential equation at $\boldsymbol{\theta}_{t-1}$. \added{We approximate $r(\boldsymbol{\theta}_{t-1}; \alpha, \beta, \gamma, f_{x_t})$ according to the 4-th order Runge-Kutta (RK4) method \citep{kutta1901beitrag}. The details of RK4 approximation are described in Section \ref{sec: RK4 details}.} An alternative method to approximate the solution to an ordinary differential equation is Euler's method \citep{butcher2000numerical}. However, RK4 has been found to be more stable, efficient, and accurate compared to Euler's method \citep{islam2015comparative, kamruzzaman2018comparative}.

Regarding to the observation equation, we denote $y_t \in [0,1]$ as the observed infectious proportion at time $t$. As $y_t$ is the observed value of $I_t$, it is not necessarily equal to $I_t$ because not all of the infectious population can be identified through diagnosis. The proportion of confirmed cases among the infected population is denoted by $p \ (\text{where }~0 < p \le 1)$, which is called the \textit{identification rate}. We assume that the observed infectious proportion $y_t$ follows a Beta distribution 
\begin{equation}
    \begin{aligned}
    \label{eq: BDSSS-SEIR-Observation Equation}
    y_t &|\boldsymbol{\theta}_t \text{\added{, $\psi$}} \sim \text{Beta}(\lambda p I_t, \lambda (1- p I_t)),
    \end{aligned}
\end{equation}
where $\lambda$ is a scaling parameter that controls the variability of the observed proportions, and $pI_t$ is the expected observed infectious proportion at time $t$. 

From herein, we delve into the mathematical details to gain a comprehensive understanding of the dynamic model. The unknown parameters of this model are $\psi = [\alpha, \beta, \gamma, \lambda, \kappa, p, [\pi_{ij}]_{K \times (K-1)}, f_2, \ldots, f_K]$. The observation process is governed by the Beta distribution in (\ref{eq: BDSSS-SEIR-Observation Equation}), and the state process is governed by the Dirichlet distribution in (\ref{eq: BDSSS-SEIR-Transition Equation}). Notably, the conditional expectation and variance of the observed proportion of positive cases $y_t$ are
\begin{align*}
E(y_t|\boldsymbol{\theta}_t, \psi) & =  p  I_t ,\\
Var(y_t|\boldsymbol{\theta}_t, \psi) & = \frac{ p  I_t(1-  p I_t)}{\lambda + 1}.
\end{align*}
In particular, the expected value $E(y_t|\boldsymbol{\theta}_t, \psi)$ represents the average proportion of infectious population that is identified as positive at time $t$. It is biased with respect to the latent state $I_t$ due to the identification rate $p$. The variance $Var(y_t|\boldsymbol{\theta}_t, \psi)$ quantifies the variability in the observed proportion of positive cases $y_t$ at time $t$, given the current state of the system and the model parameters. This variance depends on $p$, $I_t$, and the scaling parameter $\lambda$. The term $\frac{p I_t (1-p I_t)}{\lambda + 1}$ reflects the fact that the variance increases as $p$ and $I_t$ increase, while being restrained by the precision parameter $\lambda$, which can be adjusted to match the observed data more closely. Furthermore, according to the transition equation in (\ref{eq: BDSSS-SEIR-Transition Equation}), the conditional expectation and variance of $\boldsymbol{\theta}_t$ are
\begin{align*}
E(\boldsymbol{\theta}_t|\boldsymbol{\theta}_{t-1}, \psi) = [\eta_t^S, \eta_t^E, \eta_t^I, \eta_t^R]^\top, \\
\begin{bmatrix}
Var(S_t|\boldsymbol{\theta}_{t-1}, \psi)\\
Var(E_t|\boldsymbol{\theta}_{t-1}, \psi)\\
Var(I_t|\boldsymbol{\theta}_{t-1}, \psi)\\
Var(R_t|\boldsymbol{\theta}_{t-1}, \psi)
\end{bmatrix}
=
\begin{bmatrix}
\frac{\eta_t^S (1-\eta_t^S)}{1+\kappa} \\ \frac{\eta_t^E (1-\eta_t^E)}{1+\kappa} \\
\frac{\eta_t^I (1-\eta_t^I)}{1+\kappa} \\ \frac{\eta_t^R (1-\eta_t^R)}{1+\kappa}
\end{bmatrix}.
\end{align*}
The parameterization of the Dirichlet distribution allows us to embed the SEIR model into the conditional mean structure of the transition equation while regulating the conditional variance using the parameter $\kappa$. The parameter $\kappa$ is added to the denominator of the variance terms for each SEIR compartment to account for overdispersion. As the value of $\kappa$ increases, the variance of the potential susceptible, exposed, infected, and recovered proportion decreases. Thus, $\kappa$ plays a key role in determining the accuracy of estimating latent trajectory $\boldsymbol{\theta}_{1:T}$.

\begin{figure}[ht]
    \centering
  \includegraphics[scale=0.65]{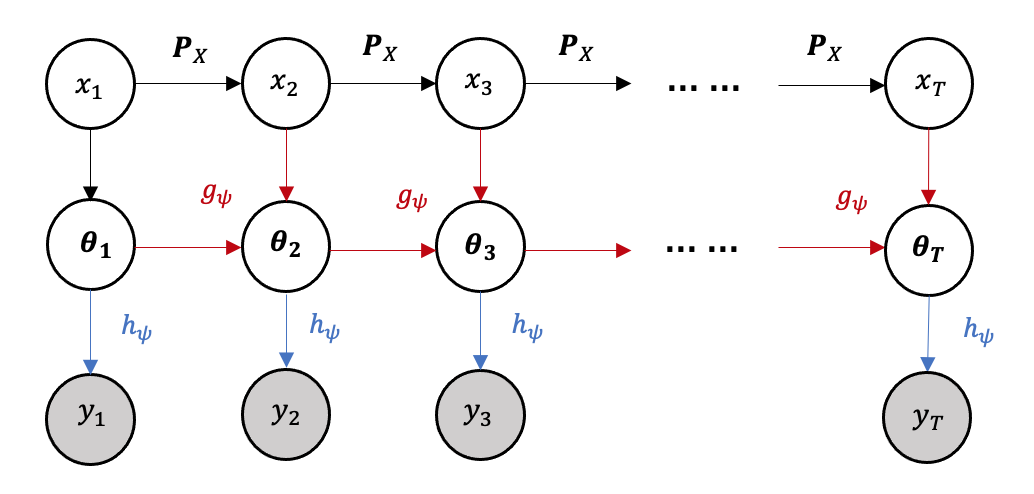}
 \caption{Schematic diagram of the switching state-space model. Circles in white indicate ``unobserved'', while circles in grey indicate ``observed''. Assume the initial latent variables $x_1$ and $\boldsymbol{\theta}_1$ are generated from $\mu_\psi(x_1)$ and $\mu_{\psi}(\boldsymbol{\theta}_1|x_1)$ respectively. The observation process $y_{1:T}$ and latent state process $\boldsymbol{\theta}_{1:T}$ are characterized by probability distributions $h_\psi(\cdot)$ and $g_\psi(\cdot)$ under the hidden Markov process $x_{1;T}$ governed by the transition probability matrix $\boldsymbol{P}_X$. }
    \label{fig:switching state-space flow chart}
\end{figure}

In general, a switching state-space model extends a state-space model by introducing a switching mechanism that governs the model's status over time \citep{shumway1991dynamic, kim1999state,deng2004switching,  costa2006discrete, nonejad2015particle, karame2018new, kim2017permanent, taghia2017bayesian, degras2022markov}. In our BDSSS-SEIR model, the switching mechanism is implemented through the transition equation, which allows for the representation of different transmission regimes over time. These regimes can capture the timing and magnitude of variations in transmission rates, reflecting the effectiveness of interventions. To better understand our switching state-space model, we provide a schematic diagram to describe the dependence structure of our BDSSS-SEIR model in Figure \ref{fig:switching state-space flow chart}. For simplicity, we use $\boldsymbol{P}_X = [\pi_{ij}]_{i,j \in \{1,\ldots,K\}}$ to denote the transition probability matrix that drives the hidden Markov process $x_{1:T}$. Conditional upon $\psi$ and $x_{1:T}$, the sequence of observations $y_{1:T}$ is modeled by a probabilistic relationship between observation process $y_{1:T}$ and latent state process $\boldsymbol{\theta}_{1:T}$, denoted as $h_\psi(\cdot)$, and a probabilistic relationship between latent state process $\boldsymbol{\theta}_{1:T}$ and the switching state process $x_{1:T}$ , denoted as $g_\psi(\cdot)$. We assume the presence of a Markovian property within the process.  It is important to note that in a more generalized case, the observation process $h_\psi(\cdot)$ and the latent state process $g_\psi(\cdot)$ can take on various forms, including discrete, continuous, or mixture distribution.

\subsection{Bayesian inference in switching state-space models}
\label{sec: Bayesian Inference}
Our primary goal is to perform Bayesian inference for the proposed BDSSS-SEIR model. The two sets of latent variables $\boldsymbol{\theta}_{1:T} = \{\boldsymbol{\theta}_1, \ldots, \boldsymbol{\theta}_T\}$ and $x_{1:T} = \{x_1, \ldots, x_T\}$ as well as the model parameters $\psi$ are treated as unknowns and
jointly estimated based on the posterior density:
\begin{equation}
    \begin{aligned}
    \label{eq:posterior distribution}
    p(\boldsymbol{\theta}_{1:T}, x_{1:T}, \psi|y_{1:T}) & \propto p_\psi(\boldsymbol{\theta}_{1:T}, x_{1:T}, y_{1:T}) \pi(\psi) \\
    & \propto p_\psi(y_{1:T}|\boldsymbol{\theta}_{1:T}, x_{1:T}) p_\psi(\boldsymbol{\theta}_{1:T}|x_{1:T}) p_\psi(x_{1:T}) \pi(\psi) \\
    & = \bigg[\prod_{t=1}^T h_\psi(y_t|\boldsymbol{\theta}_{t}, x_{t}) \prod_{t=2}^T g_\psi(\boldsymbol{\theta}_t | \boldsymbol{\theta}_{t-1}, x_{t}) \prod_{t=2}^T p_\psi(x_t|x_{t-1}) \bigg] \mu_{\psi}(\boldsymbol{\theta}_1 | x_1) \mu_{\psi}(x_1) \pi(\psi) \added{,}
\end{aligned}
\end{equation}
where $\pi(\psi)$ is the prior distribution of model parameters (see Section \ref{sec: prior on parameters} for details). We assume that the distribution of initial states $\boldsymbol{\theta}_1$ and $x_1$ are associated with the prior densities, as illustrated in Section \ref{sec: prior on latent variables}. Since the posterior distribution $p(\boldsymbol{\theta}_{1:T}, x_{1:T}, \psi|y_{1:T})$ does not have a closed-form expression, it is necessary to lean upon an efficient approximation strategy. 

There are two major reasons for using a particle MCMC method compared to traditional MCMC methods when applied to the BDSSS-SEIR model. First, the BDSSS-SEIR model is nonlinear and non-Gaussian, which can make sampling from the posterior distribution challenging. Particle MCMC can approximate the complex posterior distribution of the model more effectively than traditional MCMC methods.  Second, the strong correlation between the \added{time-varying} latent variables can potentially slow down the mixing of the Markov chain.  \added{Particle MCMC leverages the efficiency of SMC in handling time-varying latent variables, resulting in improved mixing of the Markov chain \citep{endo2019introduction}.} Hence, we use a particle Gibbs sampling approach \citep{andrieu2010particle} to explore the posterior distribution. Since conjugate priors are not available, Metropolis-Hasting steps are incorporated within the particle Gibbs sampler to draw model parameters.

This section provides an overview of the particle Gibbs sampler in the context of the BDSSS-SEIR model. We begin with SMC, which is used to initialize the reference trajectories for the entire particle Gibbs sampler. Next, we illustrate the partially deterministic conditional SMC with ancestor sampling for the joint smoothing distribution $p_\psi(\boldsymbol{\theta}_{1:T}, x_{1:T} | y_{1:T})$. Finally, we describe the particle Gibbs sampler for Bayesian learning of the proposed model. 

\subsubsection{Sequential Monte Carlo}

We start by reviewing the standard SMC, which is used to initialize the reference trajectory. The key idea of SMC \citep{doucet2001introduction} is to generate a system of weighted particles that represent possible values of the latent variables at each time step. These particles are updated sequentially over time by incorporating new observations and discarding particles with low weights. Suppose $N$ is the number of particles at a time step. Let $\{\boldsymbol{\theta}_{t-1}^{(i)}, x_{t-1}^{(i)}\}_{i=1}^N$ be the particles with normalized weights $\{W_{t-1}^{(i)}\}_{i=1}^N$ at time $t-1$. The particles and their weights are updated from time $t-1$ to time $t$ through a two-step process: importance sampling (IS) and resampling. 

First, in the IS step, the $i$-th particle is propagated based on a proposal distribution $q_\psi(\boldsymbol{\theta}_{t}^{(i)}, x_{t}^{(i)} \mid \boldsymbol{\theta}_{t-1}^{(i)}, x_{t-1}^{(i)}, y_{t})$ and assigned an importance weight $w_t^{(i)}$ 
\begin{align*}
\label{eq: unnormalized importance weight}
   w_t^{(i)}
   & \propto \frac{h_{\psi} (y_{t} \mid \boldsymbol{\theta}_{t}^{(i)}, x_{t}^{(i)}) g_{\psi}\left(\boldsymbol{\theta}_{t}^{(i)} \mid \boldsymbol{\theta}_{t-1}^{(i)}, x_{t}^{(i)}\right) p_{\psi}\left(x_{t}^{(i)} \mid x_{t-1}^{(i)}\right)}{q_\psi(\boldsymbol{\theta}_{t}^{(i)}, x_{t}^{(i)} \mid \boldsymbol{\theta}_{t-1}^{(i)}, x_{t-1}^{(i)}, y_{t})}.
\end{align*}
The importance weight represents the likelihood of obtaining a specific sample from the true posterior distribution given the proposal distribution. It is used to adjust for the discrepancy between the proposal distribution and the true posterior distribution. A detailed derivation of the importance weight is described in \ref{sec: Derivation of Importance Weights}.  In practice, it is preferable to choose a proposal distribution that is similar to the target so that a finite number of weighted particles can estimate the target distribution closely. We propagate particles by directly simulating from the state transition density in (\ref{eq: BDSSS-SEIR-Transition Equation}) and regime transition density in (\ref{eq:transitionProb}). The proposal density becomes 
\begin{equation}
\label{eq: SMC proposal density}
q_\psi(\boldsymbol{\theta}_{t}^{(i)}, x_{t}^{(i)} \mid \boldsymbol{\theta}_{t-1}^{(i)}, x_{t-1}^{(i)}, y_{t}) = g_{\psi}\left(\boldsymbol{\theta}_{t}^{(i)} \mid \boldsymbol{\theta}_{t-1}^{(i)}, x_{t}^{(i)}\right) p_{\psi}\left(x_{t}^{(i)} \mid x_{t-1}^{(i)}\right),
\end{equation}
and the importance weight simplifies to
\begin{equation}
\label{eq: simplified importance weights}
    w_t^{(i)} \propto h_\psi(y_t|\boldsymbol{\theta}_t^{(i)}, x_t^{(i)}).
\end{equation}
Once $w_t^{(i)}$'s are available, the normalized importance weights $W_t^{(i)}$ are obtained as
\begin{align}
   W_t^{(i)} = \frac{w_t^{(i)}}{\sum_{j=1}^N w_t^{(j)}}, \ \ i=1,\ldots,N,
\end{align}
where $\sum_{i=1}^N W_t^{(i)} = 1$. 

Second, in the resampling step, the weighted particles in IS step are resampled to yield equally weighted particles. We employ the multinomial resampling procedure, where particles at time $t$ choose their ancestral particles at time $t-1$ based on normalized weights $\{W_{t-1}^{(i)}\}_{i=1}^N$. This is done by sampling ancestor indices $\{a_t^{(i)}\}_{i=1}^N$ with replacement according to the probabilities $\{W_{t-1}^{(i)}\}_{i=1}^N$, where $a_t^{(i)}$ represents the index of the ancestral particle of $\{\boldsymbol{\theta}_t^{(i)}, x_t^{(i)}\}$. By concatenating the ancestral path and propagated particles at each time step, the particle trajectories can be recursively defined as $   \boldsymbol{\theta}_{1:t}^{(i)}= \{\boldsymbol{\theta}_{1:t-1}^{(a_t^{(i)})}, \boldsymbol{\theta}_t^{(i)}\}, \ x_{1:t}^{(i)} = \{x_{1:t-1}^{(a_t^{(i)})}, x_t^{(i)}\}$.
This resampling step filters out the particles with low weights and leave more informative particles in the system. The SMC procedure is initialized by sampling from a proposal density associated with priors 
\begin{equation}
\label{eq: initial SMC proposal}
q_\psi(\boldsymbol{\theta}_1, x_1|y_1) = \mu_{\psi}(x_1) \mu_\psi(\boldsymbol{\theta}_1|x_1).
\end{equation}
The specific priors of $\boldsymbol{\theta}_1$ and $x_1$ chosen to implement our proposed model are discussed in Section \ref{sec: prior on latent variables}.

Sequentially, the importance sampling and resampling steps are carried forward in time to generate particles. At terminal time $T$, the SMC sampling procedure provides us with an approximation of the joint smoothing density
\begin{equation} 
\label{eq: approximate joint smoothing density}
\hat{p}_{\psi}\left(\boldsymbol{\theta}_{1: T}, x_{1: T} \mid y_{1: T}\right)=\sum_{i=1}^{N} W_T^{(i)} \delta_{ (\boldsymbol{\theta}_{1: T}^{(i)}, x_{1: T}^{(i)} )} \left(\boldsymbol{\theta}_{1:T}, x_{1:T}\right),
\end{equation}
where $\delta_{ (\boldsymbol{\theta}_{1:T}^{(i)}, x_{1: T}^{(i)}) }\left(\boldsymbol{\theta}_{1:T}, x_{1:T}\right)$ assigns a point mass on each particle trajectory $\{\boldsymbol{\theta}_{1:T}^{(i)}, x_{1:T}^{(i)}\}_{i=1}^N$. Additionally, SMC can also provide an estimate of the likelihood of the data $\hat{p}_\psi (y_{1:T})$ using the produced particle weights. This likelihood of data is also known as the marginal likelihood because it integrates out $\boldsymbol{\theta}_{1:T}$ and $x_{1:T}$. We take advantage of this SMC property for model comparison to choose the optimal number of regimes $K$ in real data analysis. The marginal likelihood $p_\psi (y_{1:T})$ is estimated using the un-normalized importance weights $\{w_t^{(i)}\}_{i=1}^N$
\begin{equation}
\label{eq:approximate marginal likelihood}
    \hat{p}_\psi(y_{1:T})=\hat{p}_\psi(y_1) \prod_{t=2}^T \hat{p}_\psi(y_t|y_{:t-1}) = \prod_{t=1}^T \bigg[\frac{1}{N} \sum_{i=1}^N w_t^{(i)} \bigg],
\end{equation}
where 
\begin{equation}
\hat{p}_\psi(y_1) = \frac{1}{N} \sum_{i=1}^N w_1^{(i)}, \  \hat{p}_\psi (y_t|y_{1:t-1}) = \frac{1}{N}\sum_{i=1}^{N} w_t^{(i)},
\end{equation}
are the estimates of 
\begin{equation}
     p_\psi(y_1) = \int_{\boldsymbol{\theta}} \sum_{x_{1}} f_{\psi}\left(y_{1} \mid \boldsymbol{\theta}_{1}, x_{1}\right) \mu_{\psi}\left(\boldsymbol{\theta}_{1}| x_{1} \right) \mu_{\psi}\left(x_{1} \right) d \boldsymbol{\theta}_{1},
\end{equation}
\begin{equation}
    p_\psi(y_t \mid y_{1:t-1}) = \int_{\boldsymbol{\theta}} \sum_{x_{1:t}} f_\psi \left(y_t \mid \boldsymbol{\theta}_t, x_t\right) g_\psi \left(\boldsymbol{\theta}_t \mid \boldsymbol{\theta}_{t-1}, x_{t}\right) p_\psi \left(x_t \mid x_{t-1} \right) \times  p_{\psi}\left(\boldsymbol{\theta}_{1:t-1}, x_{1:t-1} \mid y_{1: t-1}\right) d \boldsymbol{\theta}_{1:t}.
\end{equation}
The SMC sampler for the proposed BDSSS-SEIR model is summarized in Algorithm \ref{algorithm: SMC}. We refer readers to Algorithm \ref{algorithm: PG-CSMC-AS} for the detailed usage of SMC in initializing reference trajectories.

\begin{algorithm}[H]
\caption{SMC for BDSSS-SEIR Model}\label{algorithm: SMC}
\hspace*{\algorithmicindent} \textbf{Input:} $y_{1:T}, N, \psi$ \\
 \hspace*{\algorithmicindent} \textbf{Output:}  $\{\boldsymbol{\theta}_{1:T}^{(i)}, x_{1:T}^{(i)}\}_{i=1}^{N}, \{W_{1:T}^{(i)}\}_{i=1}^N,\{a_{2:T}^{(i)}\}_{i=1}^N$
 \begin{algorithmic}[1]
\State \textbf{\underline{At time $t=1$}}
\State Sample $\{\boldsymbol{\theta}_{1}^{(i)}, x_{1}^{(i)}\}_{i=1}^N \sim q_\psi(\cdot, \cdot|y_1)$ according to Equation (\ref{eq: initial SMC proposal}).
\State Compute un-normalized and normalized importance weights
\begin{align*}
w_1^{(i)} = \frac{h_\psi(y_1|\boldsymbol{\theta}_1^{(i)}, x_1^{(i)}) \mu_\psi(\boldsymbol{\theta}_1^{(i)}|x_1^{(i)}) \mu_\psi(x_1^{(i)})}{q_\psi(\boldsymbol{\theta}_1^{(i)}, x_1^{(i)}|y_1)},
W_1^{(i)} = \frac{w_{1}^{(i)} }{\sum_{j=1}^N w_{1}^{(j)}}
\end{align*}
for $i=1,\ldots,N$.

\State \textbf{\underline{At time $t = 2, \ldots , T$}}
\State Sample ancestor indices $\{a_{t}^{(i)}\}_{i=1}^{N}$ with the probabilities $\{W_{t-1}^{(i)}\}_{i=1}^N$.
\State Sample $\{\boldsymbol{\theta}_t^{(i)}, x_{t}^{(i)}\}_{i=1}^{N} \sim q_\psi(\cdot, \cdot|\theta_{1:t-1}^{(a_t^{(i)})}, x_{1:t-1}^{(a_t^{(i)})}, y_t)$ according to Equation (\ref{eq: SMC proposal density}). Set $\boldsymbol{\theta}_{1:t}^{(i)}= \{\boldsymbol{\theta}_{1:t-1}^{(a_t^{(i)})}, \boldsymbol{\theta}_t^{(i)}\}$ and $x_{1:t}^{(i)} = \{x_{1:t-1}^{(a_t^{(i)})}, x_t^{(i)}\}$ for $i=1,\ldots,N$.
\State Compute un-normalized and normalized importance weights
\begin{align*}
    w_t^{(i)} = \frac{h_{\psi} (y_{t} \mid y_{1:t-1}, \boldsymbol{\theta}_{1: t}^{(i)}, x_{1: t}^{(i)}) g_{\psi}\left(\boldsymbol{\theta}_{t}^{(i)} \mid \boldsymbol{\theta}_{1: t-1}^{(a_t^{(i)})}, x_{1: t}^{(i)}\right) p_{\psi}\left(x_{t}^{(i)} \mid x_{t-1}^{(a_t^{(i)})}\right)}{q_\psi(\boldsymbol{\theta}_{t}^{(i)}, x_{t}^{(i)} \mid \boldsymbol{\theta}_{1: t-1}^{(a_t^{(i)})}, x_{1: t-1}^{(a_t^{(i)})}, y_{1:t})}, W_t^{(i)} = \frac{w_{t}^{(i)} }{\sum_{i=1}^N w_{t}^{(i)}}
\end{align*}
for $i=1,\ldots,N$.
\end{algorithmic}
\end{algorithm}


\subsubsection{Conditional Sequential Monte Carlo}
\label{sec: conditional SMC}

In a standard particle Gibbs sampler, a conditional SMC update is required to establish a valid Markov kernel \citep{andrieu2010particle}. The difference between conditional SMC and SMC lies in whether a reference trajectory $\{\boldsymbol{\theta}_{1:T}^{(B_{1:T})}, x_{1:T}^{(B_{1:T})}\}$ is pre-specified and ensured to survive through the resampling process. Here $B_{1:T}$ denotes the ancestral lineage of the trajectory that survived in the particle system. The reference trajectory is usually frozen at a fixed location over time, establishing a valid Markov kernel and leaves the target distribution invariant \citep{andrieu2010particle}. Since the distribution of the particle system does not change even if the particle labels are permuted, there is no difference between assigning $\{B_1=1, \ldots, B_T=1\}$ or $\{B_1=N, \ldots, B_T=N\}$, i.e. storing the reference trajectory at the first or last position of the particle system at each time point. The remaining $N-1$ particles are generated as in standard SMC. After a complete pass of conditional SMC, an updated reference trajectory is sampled from the estimated $p_\psi(\boldsymbol{\theta}_{1:T}, x_{1:T}|y_{1:T})$ according to path weights $\{W_T^{(i)}\}_{i=1}^N$. This reference trajectory is saved and brought to the next MCMC iteration. 

However, the standard particle Gibbs sampler often suffers from the path degeneracy problem because many past particle trajectories are discarded inevitably in resampling step, resulting in the reduced diversity of particle trajectories at each time point. As a consequence, the particle system $\{\boldsymbol{\theta}_{1:T}^{(i)}, x_{1:T}^{(i)}\}_{i=1}^N$ shares a few common ancestral paths. This phenomenon is the so-called path degeneracy. When the continuous state of the dynamic system is extended by the discrete state variable, the path degeneracy problem becomes even more severe, leading to very poor mixing of Markov kernel \citep{andrieu2003efficient, driessen2004efficient}. To address this issue, we employed a partially deterministic conditional SMC 
within the PG sampler \citep{kim2015efficient}, along with an ancestor sampling (AS) step \citep{lindsten2014particle} to improve the mixing of MCMC kernel. There are two major differences between the partially deterministic CSMC-AS and the standard CSMC. First, the discrete switching state variable is assigned values deterministically as
    $$ x_t^{(i)} = k \text{ for } i \in \{(k-1)M+1, (k-1)M+2, \ldots, kM\}, $$
for $k=1,2,\ldots,K$. Specifically, for each value of $x_t \in \{1,\ldots,K\}$, we generate $M$ particles of $\boldsymbol{\theta}_t$ conditional on the past particle trajectories up to time $t-1$. We use $N = KM$ to denote the total number of particles for notational consistency. This deterministic exploration for the parameter space of $x_t$ guarantees the number of particles to be the same for each value of $x_t$, therefore alleviates the particle degeneracy problem. The proposal distribution associated with this partially deterministic sampling approach is given as
\begin{equation}
\label{eq:Proposal Distribution}
q_\psi(\boldsymbol{\theta}_t, x_t | \boldsymbol{\theta}_{1:t-1}, x_{1:t-1}, y_t) \propto q_\psi(\boldsymbol{\theta}_t | \boldsymbol{\theta}_{1:t-1}, x_{1:t}, y_t),
\end{equation}
where we ignore the particle index for brevity. Second, the reference particle set $\{\boldsymbol{\theta}_t^{(B_t)}, x_t^{(B_t)}\}$ is stored at the $m$-th position, where $m=x_t^{(B_t)}M$. This can be considered as storing the reference particle set $\{\boldsymbol{\theta}_t^{(B_t)}, x_t^{(B_t)}\}$ at the last position of the sub-particle system that $x_t^{(B_t)}$ points to. For instance, if $x_t^{(B_t)}=2$, $\{\boldsymbol{\theta}_t^{(B_t)}, x_t^{(B_t)}\}$ is stored at the $2M$-th position of the particle system at time $t$. The AS step sequentially updates each component in the reference trajectory over time by drawing the index $a_t^{(m)}$ with AS weights
\begin{equation} \label{eq: AS weights}
    P(a_t^{(m)}=i) \propto h_{\psi}(y_t|\boldsymbol{\theta}_t^{(B_t)}, x_t^{(B_t)}) g_{\psi}(\boldsymbol{\theta}_t^{(B_t)}|\boldsymbol{\theta}_{t-1}^{(i)}, x_t^{(B_t)})p_{\psi}(x_t^{(B_t)}|x_{t-1}^{(i)}) W_{t-1}^{(i)}
\end{equation}
for $t=2,\ldots,T$. The AS step assigns a historical path $\{\boldsymbol{\theta}_{1:t-1}^{(i)}, x_{1:t-1}^{(i)}\}$ to the reference trajectory $\{\boldsymbol{\theta}_{t:T}^{(B_{t:T})}, x_{t:T}^{(B_{t:T})}\}$, enabling higher update rates for latent states, therefore improving the mixing of MCMC kernel. The CSMC-AS update for approximating $p_\psi(\boldsymbol{\theta}_{1:T}, x_{1:T}|y_{1:T})$ is outlined in Algorithm \ref{algorithm: CSMC-AS}. The estimated target density $\hat{p}_\psi (\boldsymbol{\theta}_{1:T}, x_{1:T}|y_{1:T})$ and marginal likelihood $\hat{p}_\psi(y_{1:T})$ can be obtained as illustrated in Equation (\ref{eq: approximate joint smoothing density}) and (\ref{eq:approximate marginal likelihood}). \added{Note, in CSMC-AS,  the estimator based on the product of the average weights in Equation \ref{eq:approximate marginal likelihood} is no longer unbiased because it is conditioned on a reference trajectory. But, it is still consistent, converging almost surely to  $p_\psi(y_{1:T})$ as the number of particles goes to infinity \citep{naesseth2019elements}.}

\begin{algorithm}[ht]
\caption{CSMC-AS for BDSSS-SEIR Model}\label{algorithm: CSMC-AS}
\hspace*{\algorithmicindent} \textbf{Input:}  $y_{1:T}, \{\boldsymbol{\theta}_{1:T}^{(B_{1:T})}, x_{1:T}^{(B_{1:T})} \},N=KM,  \psi$ \\
 \hspace*{\algorithmicindent} \textbf{Output:}  $\{\boldsymbol{\theta}_{1:T}^{(i)}, x_{1:T}^{(i)}\}_{i=1}^{N}, \{W_{1:T}^{(i)}\}_{i=1}^N, \{a_{2:T}^{(i)}\}_{i=1}^N$
 \begin{algorithmic}[1]
\State \textbf{\underline{At time $t=1$}}
\State Sample $\{\boldsymbol{\theta}_{1}^{(i)}, x_{1}^{(i)}\}_{i=1}^{N} \sim q_\psi(\cdot, \cdot|y_1)$ according to Equation (\ref{eq: initial SMC proposal}). Replace $\{\boldsymbol{\theta}_{1}^{(m)}, x_{1}^{(m)}\}$ with $\{\boldsymbol{\theta}_{1}^{(B_1)}, x_{1}^{(B_1)}\}$, where $m=x_1^{(B_1)}M$.
\State Compute un-normalized and normalized importance weights
\begin{align*}
w_1^{(i)} = \frac{h_\psi(y_1|\boldsymbol{\theta}_1^{(i)}, x_1^{(i)}) \mu_\psi(\boldsymbol{\theta}_1^{(i)}|x_1^{(i)}) \mu_\psi(x_1^{(i)})}{q_\psi(\boldsymbol{\theta}_1^{(i)}, x_1^{(i)}|y_1)},
W_1^{(i)} = \frac{w_{1}^{(i)} }{\sum_{j=1}^N w_{1}^{(j)}}
\end{align*}
for $i=1,\ldots,N$.
\State \textbf{\underline{At time $t = 2, \ldots , T$}}
\State Sample ancestor indices $\{a_{t}^{(i)}\}_{i=1}^{M}$ with the probability $\{W_{t-1}^{(i)}\}_{i=1}^{KM}$. Replicate $\{a_{t}^{(i)}\}_{i=1}^{M}$ for $K$ times such that
$$a_t^{(j)} = a_t^{(M+j)} = \cdots = a_t^{(kM+j)}$$
for $k=1,\ldots,K-1$ and $j=1,\ldots,M$.
\State Sample $\{\boldsymbol{\theta}_t^{(i)}, x_{t}^{(i)}\}_{i=1}^{N} \sim q_\psi(\cdot, \cdot|\boldsymbol{\theta}_{1:t-1}^{(a_t^{(i)})}, x_{1:t-1}^{(a_t^{(i)})}, y_t)$ according to Equation (\ref{eq:Proposal Distribution}). Replace $\{\boldsymbol{\theta}_{t}^{(m)}, x_{t}^{(m)}\}$ with $\{\boldsymbol{\theta}_{t}^{(B_t)}, x_{t}^{(B_t)}\}$, where $m=x_t^{(B_t)}M$.
\State Draw $a_t^{(m)}$ with probability $P(a_t^{(m)}=i)$ according to Equation (\ref{eq: AS weights}). Set $\boldsymbol{\theta}_{1:t}^{(i)}= \{\boldsymbol{\theta}_{1:t-1}^{(a_t^{(i)})}, \boldsymbol{\theta}_t^{(i)}\}$ and $ x_{1:t}^{(i)} = \{x_{1:t-1}^{(a_t^{(i)})}, x_t^{(i)}\}$ for $i=1,\ldots,N$.
\State Compute un-normalized and normalized importance weights
\begin{align*}
    w_t^{(i)} = \frac{h_{\psi} (y_{t} \mid y_{1:t-1}, \boldsymbol{\theta}_{1: t}^{(i)}, x_{1: t}^{(i)}) g_{\psi}\left(\boldsymbol{\theta}_{t}^{(i)} \mid \boldsymbol{\theta}_{1: t-1}^{(a_t^{(i)})}, x_{1: t}^{(i)}\right) p_{\psi}\left(x_{t}^{(i)} \mid x_{t-1}^{(a_t^{(i)})}\right)}{q_\psi(\boldsymbol{\theta}_{t}^{(i)}, x_{t}^{(i)} \mid \boldsymbol{\theta}_{1: t-1}^{(a_t^{(i)})}, x_{1: t-1}^{(a_t^{(i)})}, y_{1:t})}, W_t^{(i)} = \frac{w_{t}^{(i)} }{\sum_{i=1}^N w_{t}^{(i)}}
\end{align*}
for $i=1,\ldots,N$.
\end{algorithmic}
\end{algorithm}

\subsubsection{Particle Markov Chain Monte Carlo}
\label{sec: PMCMC}

With the partially deterministic CSMC-AS, the PG sampler of estimating the latent variables and unknown parameters in the proposed model can be established. As in the standard Gibbs sampler, the PG sampler iteratively draws from $p(\psi|\boldsymbol{\theta}_{1:T}, x_{1:T}, y_{1:T})$ and $p_\psi(\boldsymbol{\theta}_{1:T}, x_{1:T}|y_{1:T})$ and provides a particle approximation for $p_\psi(\boldsymbol{\theta}_{1:T}, x_{1:T}|y_{1:T})$. In each MCMC iteration, a new reference trajectory $\{ \boldsymbol{\theta}_{1:T}^{*}, x_{1:T}^{*}\}$ is sampled from the particle system by drawing the index $B_
T$ according to the importance weights $W_T^{(i)}$ at terminal time point $T$. The corresponding trajectory with index $B_T$ at time $T$ is pulled by tracing back through the ancestral lineage $\{a_{2:T}^{(i)}\}_{i=1}^{N}$. This extra step is illustrated in Algorithm \ref{algorithm: sample reference trajectory}. Sampling $\psi$ is then conducted by employing the Metropolis-Hastings (MH) algorithms conditional on the sampled reference trajectory. We provide details of MH algorithms for sampling model parameters in Section \ref{sec: MH steps}. The particle Gibbs sampler for Bayesian learning of our proposed model is summarized in Algorithm \ref{algorithm: PG-CSMC-AS}. \added{We refer readers to \cite{chopin2015particle} for the theoretical study of the particle Gibbs sampler.} \added{In practice, multiple applications of the MH updates of the parameters can be used in-between each CSMC update of the states to improve the potential efficiency of the particle Gibbs sampler. }

\added{To facilitate model selection, we estimate the marginal likelihood $p(y_{1:T})$ based on a posterior predictive approach \citep{llorente2023marginal}. We have $p(y_{1:T}) =  \int p_\psi(y_{1:T}) p(\psi|y_{1:T})d\psi$, where $p(\psi|y_{1:T})$ is the posterior distribution of $\psi$. Since  $\hat{p}_\psi(y_{1:T})$ is a consistent estimator for $p(\psi|y_{1:T})$, and samples of $\psi$ can be obtained from the PG sampler, we estimate the marginal likelihood using  $\frac{1}{R}\sum_{r=1}^R\hat{p}_{\psi[r]}(y_{1:T})$, where $\psi[r]$ is the $r$-th sample from the PG sampler after convergence.}

\begin{algorithm}[ht]
\caption{Sampling the reference trajectory}\label{algorithm: sample reference trajectory}
\hspace*{\algorithmicindent} \textbf{Input:} $\{ \boldsymbol{\theta}_{1:T}^{(i)}, x_{1:T}^{(i)}\}_{i=1}^{N},\{W_{1:T}^{(i)}\}_{i=1}^N, \{a_{2:T}^{(i)}\}_{i=1}^N$ \\
 \hspace*{\algorithmicindent} \textbf{Output:}  $\{\boldsymbol{\theta}_{1:T}^{*}, x_{1:T}^{*}\}$
\begin{algorithmic}[1]
 \State \textbf{\underline{For $t=T$}} 
 \State Draw $B_T \sim \text{Multinomial}(1;\{W_T^{(i)}\}_{i=1}^N)$. 
 \State Set $\boldsymbol{\theta}_T^{*}= \boldsymbol{\theta}_T^{(B_T)}, x_T^{*} = x_T^{(B_T)}.$
 \State \textbf{\underline{For $t=T-1,\ldots,1$}}  
 \State Obtain the index of ancestor particle $B_t=a_{t+1}^{(B_{t+1})}$.
 \State Set $\boldsymbol{\theta}_t^{*} = \boldsymbol{\theta}_t^{(B_t)}, x_t^{*} = x_t^{(B_t)}.$
\end{algorithmic}
\end{algorithm}

\begin{algorithm}[ht]
\caption{PG Kernel for BDSSS-SEIR Model}\label{algorithm: PG-CSMC-AS}
\begin{algorithmic}[1]
    \State \textbf{\underline{Initialization $r=0$}}
    \State Set $\psi[0]$ arbitrarily. 
    \State Run Algorithm \ref{algorithm: SMC} (SMC algorithm) conditional on $\psi[0]$. 
    \State Generate a reference trajectory $\{\boldsymbol{\theta}_{1:T}^{(B_{1:T})}[0], x_{1:T}^{(B_{1:T})}[0]\}$ from SMC output using Algorithm \ref{algorithm: sample reference trajectory}.

    \State \textbf{\underline{For iteration $r = 1, \ldots, R$}}
    \State Run Algorithm \ref{algorithm: CSMC-AS} (CSMC-AS algorithm) targeting $p_{\psi[r-1]}(\boldsymbol{\theta}_{1:T}|y_{1:T})$ conditional on $\{ \boldsymbol{\theta}_{1:T}^{(B_{1:T})}[r-1], x_{1:T}^{(B_{1:T})}[r-1]\}$.
    
    \State Sample a reference trajectory $\{ \boldsymbol{\theta}_{1:T}^{(B_{1:T})}[r], x_{1:T}^{(B_{1:T})}[r]\}$ from CSMC-AS output using Algorithm \ref{algorithm: sample reference trajectory}.
    
    \State Sample $\psi[r]$ from $p(\psi|\boldsymbol{\theta}_{1:T}^{(B_{1:T})}[r], x_{1:T}^{(B_{1:T})}[r], y_{1:T})$ using MH algorithms in Section \ref{sec: MH steps}.

\end{algorithmic}
\end{algorithm}

\subsection{Prior distribution on $\psi$}
\label{sec: prior on parameters}

According to the posterior distribution in (\ref{eq:posterior distribution}), the prior distribution $\pi(\psi)$ plays a critical role in Bayesian inference as it allows us to incorporate prior knowledge and beliefs about the unknown parameters into our analysis, and it can heavily influence the posterior distribution. In the BDSSS-SEIR model, it is crucial to choose informative priors to ensure accurate inference on parameters and latent trajectories. A general framework of prior distribution on these unknown parameters is specified in Table \ref{table: prior distribution}. See Section \ref{sec: posterior distribution} for a more detailed form of posterior distribution incorporating these priors.

\begin{table}[ht]
\caption{Prior distributions for BDSSS-SEIR model parameters.}
\begin{tabular}{ll}
\hline
\textbf{Parameter} & \textbf{Prior Distribution} \\
\hline
Latency rate parameter ($\alpha$) & $TN(m_\alpha, \sigma_\alpha^2, 0, + \infty), \alpha \in (0, + \infty)$  \\
\hline
Transmission rate parameter ($\beta$) & $TN(m_\beta, \sigma_\beta^2, 0, + \infty), \beta \in (0, + \infty)$ \\
\hline
Recovery rate parameter ($\gamma$) & $TN(m_\gamma, \sigma_\gamma^2, 0, + \infty), \gamma \in (0, + \infty)$  \\
\hline
Precision parameter in observation process ($\lambda$) & $\text{Gamma}(a_\lambda, b_\lambda), \lambda \in (0, +\infty)$ \\
\hline
Precision parameter in state transition process ($\kappa$) & $\text{Gamma}(a_\kappa, b_\kappa), \kappa \in (0, +\infty)$ \\
\hline
Identification rate ($p$) & $TN(m_p, \sigma_p^2, a_p, b_p), p \in (a_p, b_p)$ \\
\hline
Transition probability matrix ($\boldsymbol{P}_X$) & $\begin{bmatrix} \boldsymbol{\pi}_{1} = (\pi_{11},  \ldots, \pi_{1K}) \sim \text{Dir}(\delta_{11}, \ldots, \delta_{1K})\\ 
\boldsymbol{\pi}_{2}=(\pi_{21}, \ldots, \pi_{2K}) \sim \text{Dir}(\delta_{21},  \ldots, \delta_{2K})\\ \vdots \\ \boldsymbol{\pi}_{K}=(\pi_{K1},  \ldots, \pi_{KK}) \sim \text{Dir}(\delta_{K1}, \ldots, \delta_{KK}) \end{bmatrix}$ \\
\hline
Transmission rate modifier ($f_{x_t}$) & $\begin{cases} 1, & \text{if}\ x_t=1 \\ 
\text{Unif}(1-\frac{1}{K-1},1), & \text{if}\ x_t = 2 \\ 
\text{Unif}(1-\frac{2}{K-1},1-\frac{1}{K-1}), & \text{if}\ x_t = 3 \\ \vdots \\ 
\text{Unif}(0,\frac{1}{K-1}), & \text{if}\ x_t = K \\ 
\end{cases}$ \\
\hline
\end{tabular}
\label{table: prior distribution}
\end{table}

For epidemic parameters $\alpha, \beta$, and $\gamma$, we assume a truncated Normal distribution from the left. This indicates that these parameters are positive and tend to be close to their respective means $m_{\alpha}, m_{\beta}$, and $m_{\gamma}$. Any prior knowledge or historical information about similar diseases should be utilized while setting the hyperparameters, as suggested in previous studies \citep{dukic2012tracking, osthus2017forecasting, dehning2020inferring}. The prior variance is chosen such that it corresponds to a reasonable range of plausible values for the parameter based on prior knowledge. 

The precision parameters in Beta-Dirichlet framework govern the process error. Specifically, $\lambda$ controls the magnitude of observation error, and $\kappa$ controls the magnitude of state transition error. We follow the precedent of \cite{osthus2017forecasting} and \cite{kobayashi2020predicting} to assign Gamma distributions to $\lambda$ and $\kappa$, where
$$E(\lambda) = a_{\lambda}/b_{\lambda}, Var(\lambda) = a_{\lambda}/b_{\lambda}^2 ;$$
$$E(\kappa) = a_{\kappa}/b_{\kappa}, Var(\kappa) =  a_{\kappa}/b_{\kappa}^2.$$ 
We choose prior distributions with relatively large expected values for $\lambda$ and $\kappa$ to avoid zero values in the Beta or Dirichlet parameters, thereby maintaining numerical stability. A large $\kappa$ puts a high concentration of probability mass on each compartment, resulting in a relatively fixed proportion of individuals in each compartment. This assists in reducing the distance between $g(\theta_{t-1};\alpha, \beta, \gamma, f_{x_t})$ and $\theta_t$ because a very low Dirichlet density is obtained when they differ slightly. However, it is important to avoid setting the prior variance too large, as this may hinder the effective exploration of the target distribution by the MCMC chain. We suggest users tune with the hyperparameters based on the data on hand to ensure the convergence of MCMC.

The value of identification rate $p$ for COVID-19 has been approximated for several countries \citep{kuniya2020prediction, kobayashi2020predicting, impouma2021estimating, CDC2023}. \cite{CDC2023} estimated that 1 in 4.0 (95\,\% CI $3.4–4.7$) COVID–19 infections were reported from February 2020 to September 2021 in the United States. Based on this estimation, we set a truncated Normal distribution with mean $m_p =0.25$ and standard deviation $\sigma_p = 0.05$, lower limit $a_p = 0.1$, upper limit $b_p=0.4$, to encompass the estimated 95\,\% CI and allow for other plausible values beyond this interval in simulation studies. \added{In real data analysis, when mandatory testing is introduced or abolished during the pandemic, the identification rate would be time-varying. Multiple different identification rates need to be accounted for in this scenario. We refer readers to Section \ref{sec: real data analysis} for more details.}

Each row of the transition probability matrix  $\boldsymbol{P}_X$ is denoted by $\boldsymbol{\pi}_k$, where $k=1,\ldots,K$. The prior distribution of $\boldsymbol{\pi}_k$ is determined by a Dirichlet distribution with concentration parameters $\boldsymbol{\delta}_k$. This distribution expresses prior beliefs about the likely values of the components of $\boldsymbol{\pi}_k$, where each row has its own set of concentration parameters. In practice, frequent switching between regimes rarely occur because it takes time for interventions to take effect. The model status is more likely to remain in the same regime while switching between time points, which can be expressed using an asymmetric Dirichlet prior distribution. 

The transmission rate modifier $f_{x_t}$ is a generalized function that maps each possible switching state other than 1 to a Uniform distribution over a specific interval. The modifier assumes that the transmission rate can vary depending on the current regime of the system, and the prior reflects this uncertainty by allowing for different ranges of possible values at each regime. For example, in a system with two regimes ($K=2$), the transmission rate is fixed at 1 when $x_t = 1$. However, when $x_t = 2$, the transmission rate is drawn from a Uniform distribution with lower bound of $1-\frac{1}{K-1}=0$ and upper bound of 1, indicating that the transmission rate is less certain and can vary between 0 and 1. Similarly, when $K=3$, the prior distribution of the transmission rate modifier becomes $f_1=1$, $f_2 \sim \text{Unif}(0.5, 1)$, and $f_3 \sim \text{Unif}(0, 0.5)$. 

In general, non-informative priors are often too vague to provide useful information for the dynamic model about tracking epidemics. However, if there is expert knowledge available in the field, it is advisable to choose prior distributions as much informative as possible, and complement the rest with non-informative priors \citep{dukic2012tracking, osthus2017forecasting, dehning2020inferring, wang2020epidemiological}. Incorporating such knowledge into the model can improve the scientific rigor of the study.

\subsection{Prior distribution on $\boldsymbol{\theta}_1$ and $x_1$}
\label{sec: prior on latent variables}

The initial values of $\boldsymbol{\theta}_1$ and $x_1$ are associated with their priors throughout this paper. Assuming $\boldsymbol{\theta}_1$ and $x_1$ are independent, the joint prior density of $\boldsymbol{\theta}_1$ and $x_1$ is decomposed as
\begin{equation}
\label{eq:prior density on theta_1 and x_1}
    \mu_\psi(\boldsymbol{\theta}_1, x_1) = \mu_\psi(\boldsymbol{\theta}_1| x_1) \mu_\psi(x_1) =  \mu_\psi(\boldsymbol{\theta}_1)  \mu_\psi(x_1).
\end{equation}
Unlike existing studies that assign fixed values to the initial compartment \citep{osthus2017forecasting, kobayashi2020predicting}, we introduce some randomness to all initial compartments. The prior distribution of $\theta_1$ is a Dirichlet distribution with the mass highly concentrated in susceptible proportion:
\begin{equation}
\label{eq:prior density on theta_1 }
    \boldsymbol{\theta}_1 \sim \text{Dirichlet}(100,1,1,1).
\end{equation}
The prior distribution for $x_1$ is uniform over $K$ categories
\begin{equation}
\label{eq:prior density on x_1 }
    P(x_1 = k) = 1/K \text{ for } k=1,\ldots,K,
\end{equation}
reflecting little prior knowledge about the system regimes.

\section{Numerical results}
\label{section: Results}

\subsection{Simulation study}
\label{sec: simulation study}
To demonstrate the estimation capabilities of the particle MCMC algorithm on the model parameters $\psi$ and the latent state variables $\boldsymbol{\theta}_{1:T}$ and $x_{1:T}$, we conducted simulations in two different settings: a two-regime setting and a three-regime setting. In real-world epidemics, changes in the transmission rate may affect the dynamics of infectious population. For instance, when the transmission rate is higher in one regime than the other, switching to the high transmission rate regime may lead to an increase in the infectious proportion. By incorporating multiple waves of observed infectious proportion in our simulation study, the BDSSS-SEIR model is able to capture the impacts of external interventions on the spread of epidemics effectively, making it a valuable tool to study complicated epidemic scenarios.

\subsubsection{Two-regime setting}

\begin{figure}[ht]
\centering
\includegraphics[scale=0.2]{ 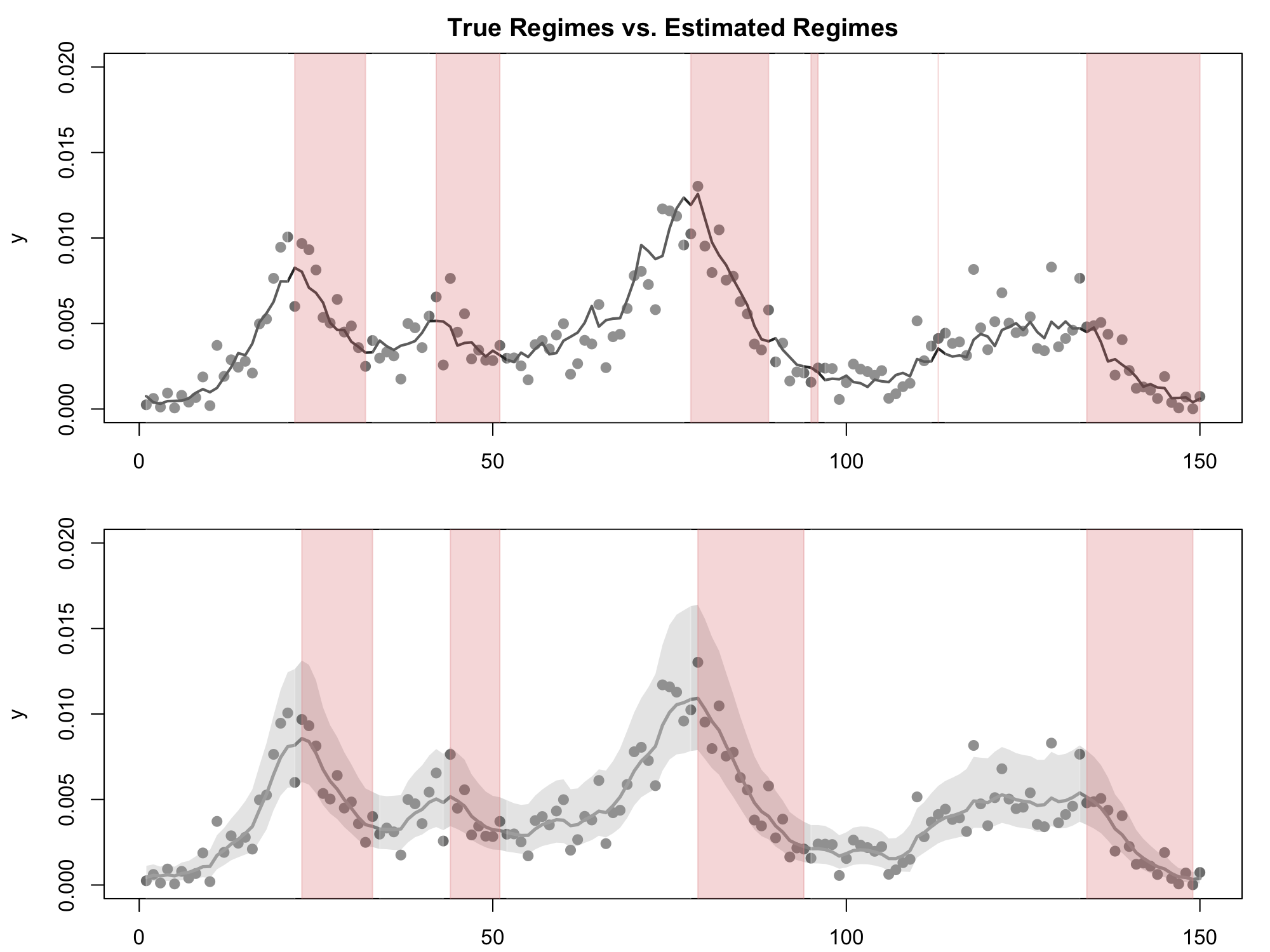}
\caption{Simulated and posterior estimates of switching states over time in the two-regime setting. The first regime ($X_t=1$) is represented by a white background, while the second regime ($X_t=2$) is indicated by a red background. Observed $y_{1:T}$ is illustrated in grey dots. In the top panel, true regimes under simulated data are shown, with $E(y_t|\boldsymbol{\theta}_t, \psi)$ 
depicted by a solid black curve. The bottom panel displays 
estimated regimes based on posterior probabilities $\hat{P}(X_t = k | y_{1:T})$ for $k=1,2$. The solid grey line with shaded area indicates the posterior mean and 95\,\% credible interval of $\hat{E}(y_{t} | \boldsymbol{\theta}_t, \psi)$ over time.}
\label{fig: True vs. Estimated Xt - two regime}
\end{figure}

We simulated the observed infectious proportion of $T=150$ time points for a two-regime setting with the transition probabilities 
$$ \boldsymbol{P}_X = \begin{bmatrix} 0.9 & 0.1 \\ 0.1 & 0.9 \end{bmatrix} \added{,} $$ 
and the transmission rate modifier 
$$f_{x_t}=\begin{cases}
      1, & \text{if}\ x_t=1, \\
      0.1, & \text{if}\ x_t = 2. 
    \end{cases}$$
The parameters in the SEIR system were set as $\alpha = 1/3, \beta=0.39,$ and $\gamma = 0.18$. In this scenario, the basic reproductive number is $\beta/\gamma = 2.17$, indicating that a single infected individual will generate 2.17 secondary infections in a completely susceptible population on average. This value falls in the range 1.4--6.49 in studies published from January 1, 2020 to February 7, 2020 \citep{liu2020reproductive}. Individuals who are exposed to the disease will get infectious in $1/\alpha=3$ days on average, while the average duration of the infectious period until recovery is $1/\gamma=5.56$ days. The precision parameters for the observation and state transition process are $\lambda=2500$ and $\kappa=5500$. The simulated dynamics of the latent variable $\boldsymbol{\theta}_{t} = [S_t, E_t, I_t, R_t]^\top$ started with initial values $\boldsymbol{\theta}_1 = [0.99, 0.001, 0.003, 0.006]^\top$. We assume the initial switching state to be $x_1=1$. The top panel of Figure \ref{fig: True vs. Estimated Xt - two regime} displays the observed infectious proportion $y_{1:T}$ drawn using the simulated $I_t$ and an identification rate $p=0.25$. The red blocks indicate the changes in regimes, resulting in four peaks in the observed infectious proportion. Noisy changes of regimes (occurs at $t=95, 96, 113$) do not impact the epidemic curve as much as that from wider blocks.

\begin{figure}[ht!]
\centering
\includegraphics[scale=0.12]{ 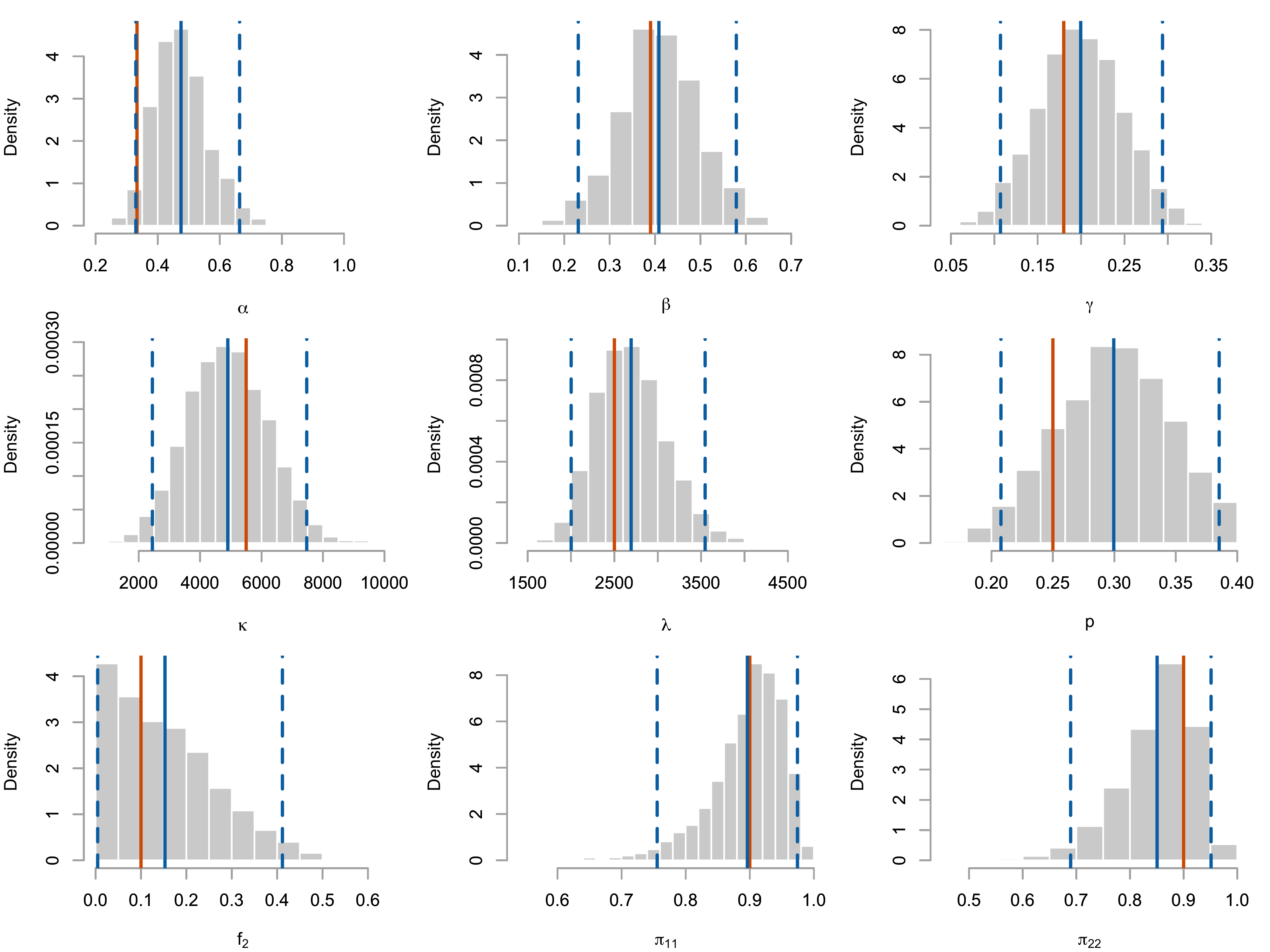}
\caption{Posterior densities for estimated model parameters. True values are indicated by vertical red lines, while the mean of the posterior densities and 95\,\% credible intervals are shown in solid and dashed blue lines.}
\label{fig:two-regime parameters histogram}
\end{figure}

The prior distributions of the model parameters $\psi$ are specified in Table \ref{table: two-regime simulation priors}. We assume the hyperparameters of $\alpha, \beta$ and $\gamma$ were derived from the historical information on similar epidemics. The transition probability matrix $\boldsymbol{P}_X$ indicates a higher likelihood of remaining in the same regime rather than switching between regimes frequently. The precision parameters $\lambda$ and $\kappa$ are assigned a Gamma distribution with $E(\lambda)=2 \times 10^3, Var(\lambda)=2 \times 10^6$ and $E(\kappa) = 2 \times 10^4, Var(\kappa) = 2 \times 10^6$, as suggested by \cite{osthus2017forecasting} and \cite{kobayashi2020predicting}. We ran \added{2 chains with} \added{30000} MCMC iterations each, discarding the first 1000 iterations of each chain. The number of particles used in each cycle of SMC or CSMC is $N=MK=100$. The total running time for this two-regime setting on a single 3 GHz Intel i5 Core is around \added{12} hours. Metropolis-Hastings steps within the particle Gibbs sampler are implemented to draw model parameters. \added{We applied five Metropolis-Hastings updates of the parameters in-between each CSMC update of the latent states.} Step sizes were tuned to guarantee an acceptance rate greater than \added{30\,\%}. The trace plots \added{and kernel density plots} for parameters are shown in Figure \ref{fig: two-regime simulation trace plot} to monitor the convergence of the particle Gibbs sampler, suggesting convergence to a stationary distribution. \added{The Gelman-Rubin diagnostic less than 1.2 \citep{brooks1998general} implies that no non-convergence issues are detected. See Table \ref{table: GR for two-regime simulation study} for more details.} The marginal posterior distributions of model parameters $\psi$ are explicitly shown in Figure \ref{fig:two-regime parameters histogram}. We propose $\pi_{11}$ and $\pi_{22}$ from a Normal distribution truncated between 0 and 1 in the MH step. The remaining transition probabilities are computed by subtracting one from the proposed element. All of the true parameter values fall within their 95\,\% Bayesian credible intervals (CI). However, we found the precision parameter $\kappa$ sometimes could be difficult to estimate. A larger prior variance of $\kappa$ can result in a wider range of possible parameter values, which can lead to a more diffuse posterior distribution, and can affect the convergence of the MCMC algorithm. The estimation of $\kappa$ also heavily relies on the reference trajectory in each MCMC iteration. We suggest users to tune the priors of $\kappa$ or choose priors on $\alpha, \beta$ and $\gamma$ as informative as possible for a more accurate estimate of $\kappa$. 

\begin{table}[H]
\centering
\caption{Summary of the true parameters and prior distribution for the two-regime data with a length of $T=150$ and an identification rate $p=0.25$.}
\begin{tabular}{lll}
\hline
\textbf{Parameter} & \textbf{Prior Distribution} & \textbf{Support} \\
\hline
$\alpha$ & $TN(0.3, 0.1^2, 0, +\infty)$ & $(0, +\infty)$ \\
\hline
$\beta$ & $TN(0.4, 0.1^2, 0, +\infty)$ & $(0, +\infty)$ \\
\hline
$\gamma$ & $TN(0.2, 0.1^2, 0, +\infty)$ & $(0, +\infty)$ \\
\hline
$\lambda$ & $\text{Gamma}(2, 0.001)$ & $(0, +\infty)$ \\
\hline
$\kappa$ & $\text{Gamma}(200, 0.01)$ & $(0, +\infty)$ \\
\hline
$p$ & $TN(0.25, 0.05^2, 0.1, 0.4)$ & $(0.1, 0.4)$ \\
\hline
$\boldsymbol{P}_X$ & $\begin{bmatrix}
\boldsymbol{\pi}_{1} \sim \text{Dir}(10, 1)\\
\boldsymbol{\pi}_{2} \sim \text{Dir}(1, 10)\\
\end{bmatrix}$ & $\boldsymbol{\pi}_{1}, \boldsymbol{\pi}_{2} \in [0,1]^2$\\
\hline
$f_{x_t}$ & $\begin{cases}
f_1 = 1, & \text{if}\ x_t=1 \\
f_2 \sim \text{Unif}(0,1),& \text{if}\ x_t=2
\end{cases}$ & $f_2 \in (0,1)$ \\
\hline
\end{tabular}

\label{table: two-regime simulation priors}
\end{table}

\begin{figure}[ht]
\centering
\includegraphics[scale=0.2]{ 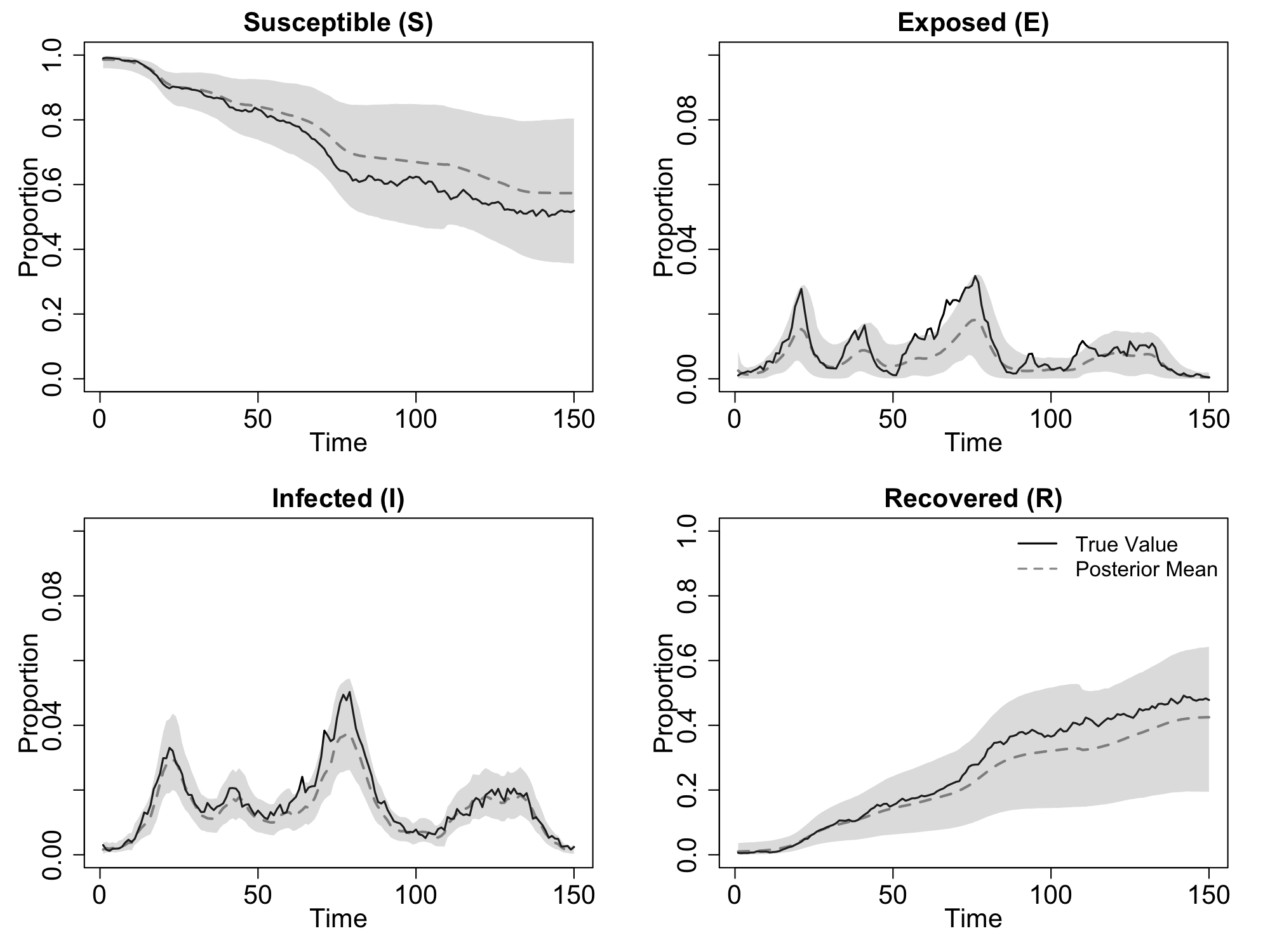}
\caption{Posterior estimates of SEIR dynamics in the two-regime setting. Simulated SEIR dynamics are represented by black lines, while the posterior means with 95\,\% credible intervals are depicted by dashed grey lines and shaded area. }
\label{fig:two-regime estimated SEIR}
\end{figure}

In Figure \ref{fig:two-regime estimated SEIR}, the 95\,\% Bayesian credible intervals of $\boldsymbol{\theta}_t = [S_t, E_t, I_t, R_t]^\top$ track the true latent trajectory effectively. As time progresses, the credible intervals widen for $S_t$ and $R_t$, indicating an increase of uncertainty in the estimates. Conversely, the credible intervals for $E_t$ and $I_t$ remain relatively constant over time, with the latter having a tighter interval.  Figure \ref{fig: True vs. Estimated Xt - two regime} displays the true and estimated regimes over time. The regimes at time $t$ is determined by the conditional posterior probability $\hat{P}(X_t|y_{1:T})$. When $\hat{P}(X_t=2|y_{1:T}) > \hat{P}(X_t=1|y_{1:T})$, the status of the model is estimated to be in the second regime at time $t$. The model's ability to accurately detect regime switching is remarkable, with a slight deviation only in the presence of noisy switches. The high-transmission-rate and low-transmission-rate regimes are precisely identified, enabling the tracking of intervention effects. A decrease in the transmission rate modifier, which could be following an intervention or vaccination program, suggests the corresponding measure's effectiveness in reducing the transmission rate. This is a valuable insight for policymakers and public health officials. An increase in the transmission rate modifier could follow an ease of external intervention, which causes the spread of disease again.

\subsubsection{Three-regime setting}

\begin{figure}[ht!]
    \centering
    \includegraphics[scale=0.2]{ 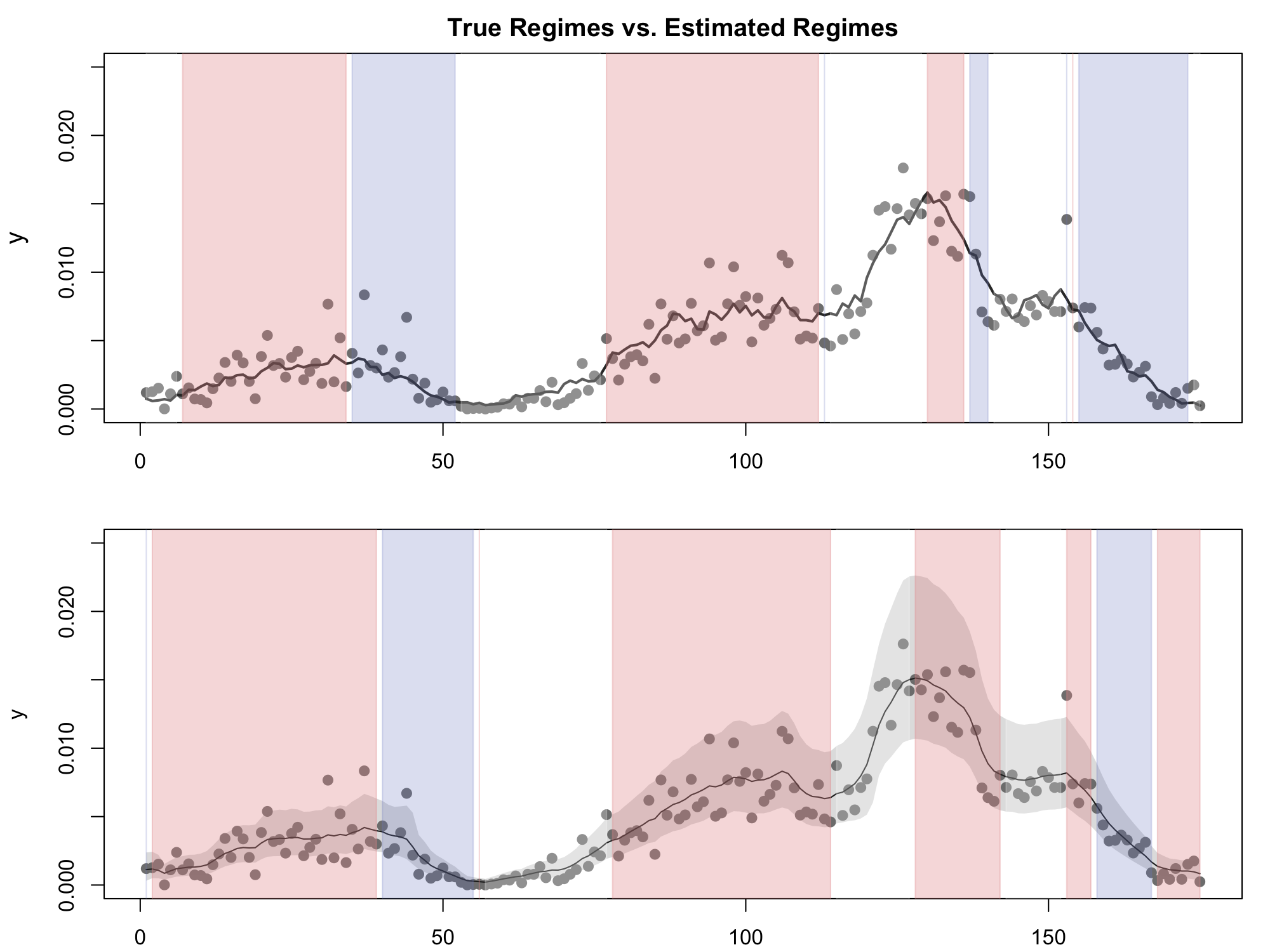}
    \caption{Simulated and posterior estimates of switching states over time in the three-regime setting. The first regime ($X_t = 1$) is indicated by a white background, the second regime ($X_t = 2$)  by a red background, and the third regime ($X_t = 3$)   by a blue background. Observed $y_{1:T}$ is represented by grey dots. (Top) True regimes under simulated data, with $E(y_{t}|\boldsymbol{\theta}_t, \psi)$ shown as a solid black curve. (Bottom) Estimated regimes based on posterior probabilities $\hat{P}(X_t = k | y_{1:T})$ for $k=1,2, 3$. The solid grey line with shaded area indicates the posterior mean and 95\,\% credible interval of $\hat{E}(y_{t}|\boldsymbol{\theta}_t, \psi)$ over time.}
    \label{fig: three regime estimated Xt}
\end{figure}

Now we move on to a more complicated scenario that allows transmission rate switches among three regimes. We simulated data of length $T=175$ with $\alpha=0.3, \beta=0.5, \gamma=0.2, \lambda=2000, \kappa=8000, \text{ and } p=0.25$. Note that the data is required to be longer for an accurate estimation in a more complicated scenario. The transition probability is extended to a $3 \times 3$ matrix
$$\boldsymbol{P}_X = \begin{bmatrix}
    0.94 & 0.03 & 0.03 \\
    0.03 & 0.94 & 0.03 \\
    0.03 & 0.03 & 0.94
\end{bmatrix} \added{,}$$
and the transmission rate modifier is defined as
$$f_{x_t} = \begin{cases}
      f_1=1, & \text{if}\ x_t=1,\\
      f_2=0.6\added{,} & \text{if}\ x_t=2,\\
      f_3=0.05\added{,} & \text{if}\ x_t=3,
    \end{cases}$$
where $x_t = 1$ indicates no external interventions occur and the transmission rate remains at its baseline value, $x_t = 2$ suggests that there are some external interventions in place that reduce the transmission rate by 40\,\%, and $x_t = 3$ indicates that there are more strict external interventions in place that reduce the transmission rate by 95\,\%. The simulated dynamics of the latent variable $\boldsymbol{\theta}_{t} = [S_t, E_t, I_t, R_t]^\top$ starts from the initial values $\boldsymbol{\theta}_1 = [0.99, 0.001, 0.003, 0.006]^\top$. The simulated $y_{1:T}$ and $x_{1:T}$ is illustrated at the top of Figure \ref{fig: three regime estimated Xt}. The reduction of transmission rate flattens the curve to some extent. The prior distributions of model parameters are specified in Table \ref{table: three-regime simulation priors}.

\begin{table}[ht]
\centering
\caption{Summary of true parameters and prior distribution for three-regime data of length $T=175$ with an identification rate $p=0.25$.}
\begin{tabular}{lll}
\hline
\textbf{Parameter} & \textbf{Prior Distribution} & \textbf{Support} \\
\hline
$\alpha$ & $TN(0.3, 0.1^2, 0, +\infty)$ & $(0, +\infty)$ \\
\hline
$\beta$ & $TN(0.4, 0.1^2, 0, +\infty)$ & $(0, +\infty)$ \\
\hline
$\gamma$ & $TN(0.2, 0.1^2, 0, +\infty)$ & $(0, +\infty)$ \\
\hline
$\lambda$ & $\text{Gamma}(20, 0.01)$ & $(0, +\infty)$ \\
\hline
$\kappa$ & $\text{Gamma}(200, 0.01)$ & $(0, +\infty)$ \\
\hline
$p$ & $TN(0.25, 0.05^2, 0.1, 0.4)$ & $(0.1, 0.4)$ \\
\hline
$\boldsymbol{P}_X$ & $\begin{bmatrix}
\boldsymbol{\pi}_{1} \sim \text{Dir}(10, 1, 1)\\
\boldsymbol{\pi}_{2} \sim \text{Dir}(1, 10, 1)\\
\boldsymbol{\pi}_{3} \sim \text{Dir}(1, 1, 10)
\end{bmatrix}$ & $\boldsymbol{\pi}_{1}, \boldsymbol{\pi}_{2}, \boldsymbol{\pi}_{3} \in [0,1]^3$\\
\hline
$f_{x_t}$ & $\begin{cases}
f_1 = 1, & \text{if}\ x_t=1 \\
f_2 \sim \text{Unif}(0.5,1),& \text{if}\ x_t=2\\
f_3 \sim \text{Unif}(0,0.5),& \text{if}\ x_t=3
\end{cases}$ & $f_2 \in (0.5,1), \ f_3 \in (0,0.5)$ \\
\hline
\end{tabular}
\label{table: three-regime simulation priors}
\end{table}

\begin{figure}[ht!]
\centering
\includegraphics[scale=0.088]{ 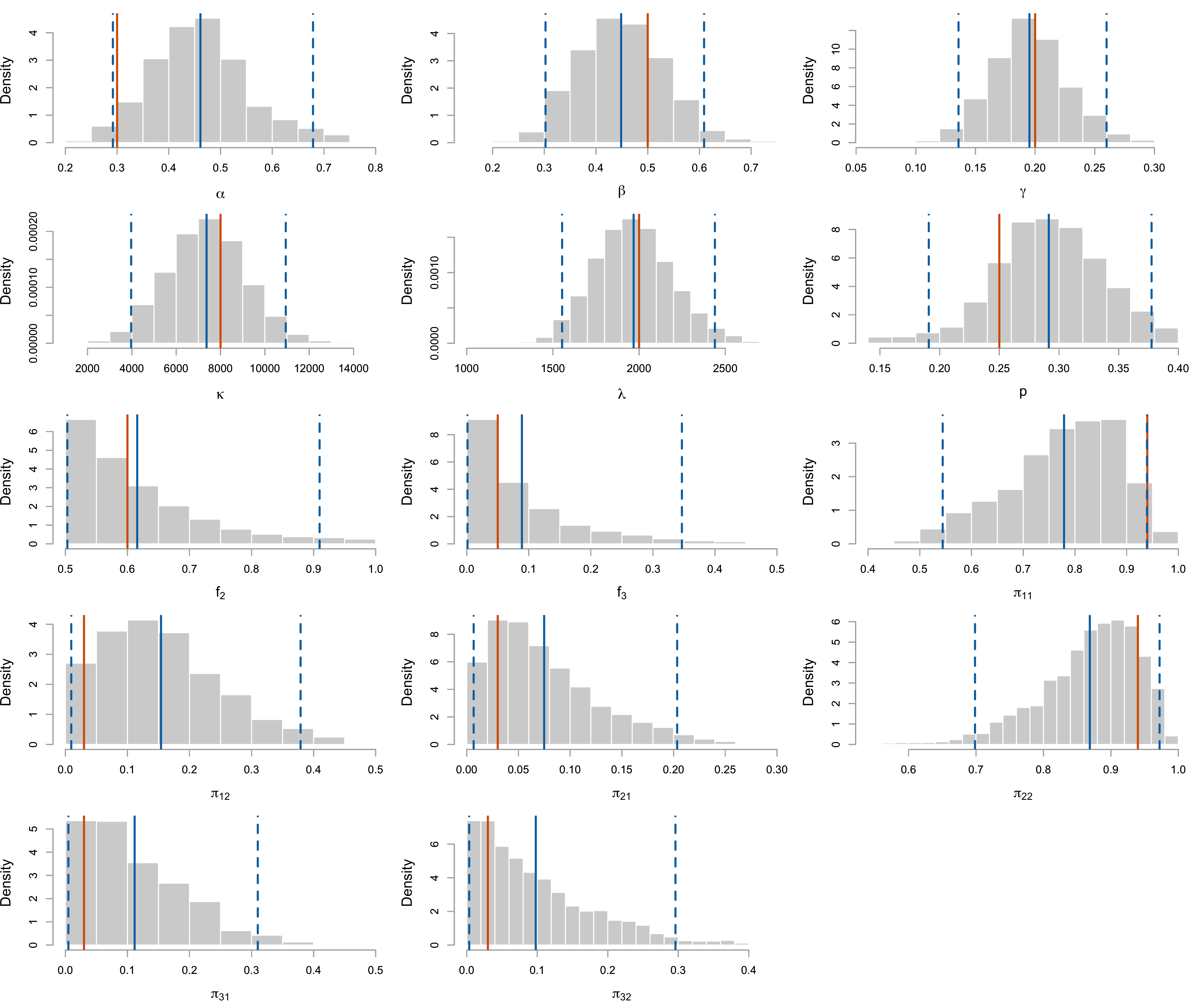}
\caption{Posterior densities for estimated model parameters in the three-regime setting. True values are indicated by vertical red lines. Mean of the posterior densities and 95\,\% credible intervals are shown in solid and dashed blue lines.}
\label{fig: three-regime simulation histogram}
\end{figure}

\begin{figure}[ht!]
    \centering 
    \includegraphics[scale=0.2]{ 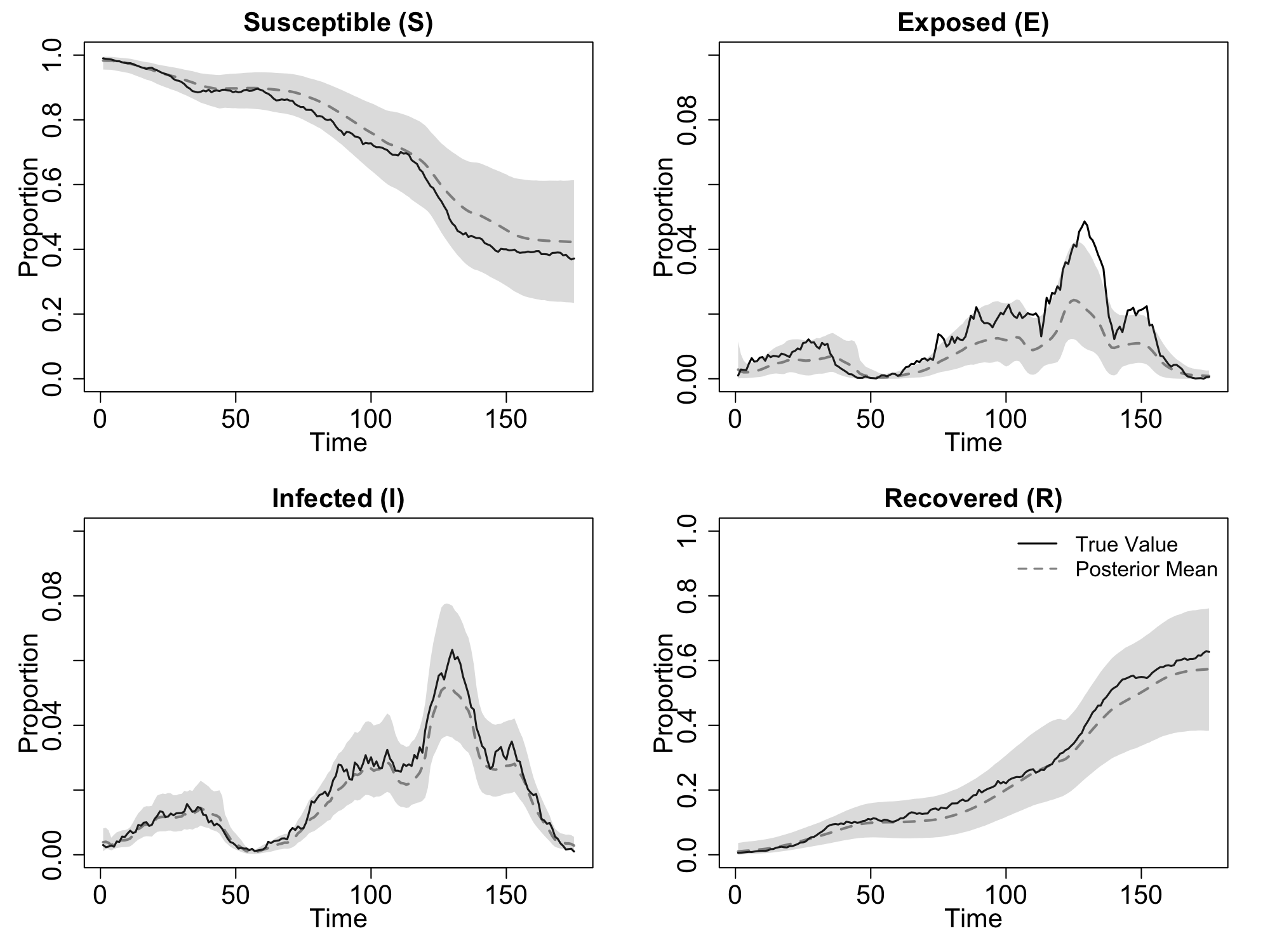}
    \caption{Posterior estimates of SEIR dynamics in three-regime setting. Simulated SEIR dynamics are represented by black lines, while the posterior mean and 95\,\% credible intervals are depicted by dashed grey lines with shaded areas. }
    \label{fig: three-regime simulation - estimated seir}
\end{figure}

We ran \added{30000} MCMC iterations with $N=MK=300$ particles after a burn-in of 1000 iterations \added{for two chains}. \added{Five Metropolis-Hastings updates for the parameters were interleaved between each CSMC update of the latent states.} The trace plots \added{and kernel density plots} of parameters shown in Figure \ref{fig: three-regime simulation trace plot} monitor the convergence of MCMC chains. The two Markov chains seem to converge to the same stationary distribution. \added{The resulting Gelman-Rubin statistic from these two chains are all less than 1.2, indicating no non-convergence issues. See more details in Table \ref{table: GR for three-regime simulation study}.} The marginal posterior distribution of parameters is displayed in Figure \ref{fig: three-regime simulation histogram}. All of the true parameter values fall within the 95\,\% credible interval, suggesting an accurate inference of parameters.
Obtaining more accurate estimates of switching states may require a larger amount of observed data. In Figure \ref{fig: three-regime simulation - estimated seir}, the series of posterior densities for the latent
variable $\boldsymbol{\theta}_t$ over time show that our model, combined with particle MCMC, can effectively track past susceptible, infected, and recovered proportions. There seems to be a higher level of uncertainty in estimating the peak of the exposed proportion compared to other variables. The results of the estimated $x_{1:T}$ are shown in Figure \ref{fig: three regime estimated Xt}. We have detected the changes in regimes, which closely match the actual regimes. The detection of switching between the second and third regime appears noisy in the latter period. Gathering more data or employing stronger priors may be necessary to make more informed inferences about the latent switching states, especially when considering more regimes.

\clearpage

\subsection{Analysis of COVID-19 data}
\label{sec: real data analysis}

In this section, we present the results of influenza tracking and regime detection using the COVID-19 Data from British Columbia, Canada. The daily active case counts are publicly reported between Jan 28, 2020 and Feb 9, 2022 in British Columbia. The corresponding data file was compiled by Hannah James, Leithen M'Gonigle, Eully Ao, and Sally Otto based on news releases from the BC Centre for Disease Control (BCCDC) and the Public Health authority \citep{datafile}. The observed proportion of infectious population is computed based on 5.07 million population in British Columbia in 2019 (Retrieved March 19, 2023, from the Data Commons website \citep{datacommons}).

\begin{figure}[ht]
    \centering
    \includegraphics[scale=0.6]{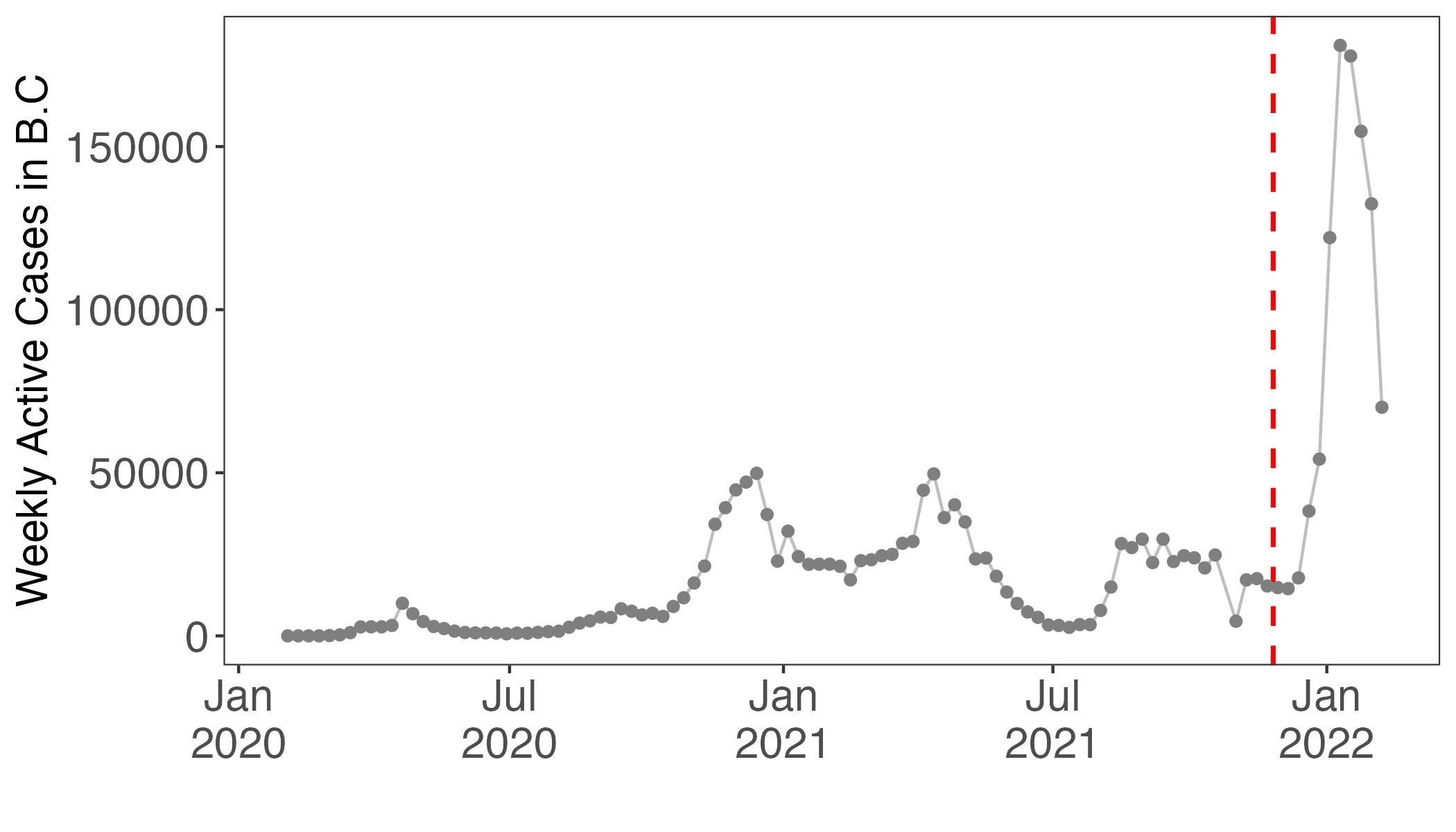}
    \caption{Weekly active case counts in British Columbia, Canada from the week of Jan 27, 2020 to the week of Feb 7, 2022 \citep{datafile}. BC identified the first case of the COVID-19 Omicron variant on Nov 30, 2021, as shown by the red dashed line \citep{cbcnews}. The observed data is of length $T=95$. }
    \label{fig:weekly active case counts in BC}
\end{figure}

The original strain of the COVID-19 virus, known as SARS-CoV-2, was first identified in Wuhan, China in late 2019. The SARS-CoV-2 virus has continued to evolve, leading to the emergence of new variants with different mutations. The Omicron variant of the SARS-CoV-2 virus is the result of several mutations that have accumulated in the virus's genome with a stronger transmissibility. The Omicron variant has more than 30 mutations in the spike protein, which is the part of the virus that allows it to infect human cells \citep{hui2022sars}. The Omicron variant (B.1.1.529) was first identified in South Africa, and it sparked a travel ban to Canada in November 2021 \citep{covid2021sars}. As seen in Figure \ref{fig:weekly active case counts in BC}, the first case of the COVID-19 Omicron variant was confirmed on Nov 30, 2021 in British Columbia. It caused an explosion of active cases soon.

However, the BDSSS-SEIR model we presented is not designed to detect the emergence of new variants of the virus because researchers and public health officials typically rely on genomic sequencing to detect the new variant of the virus. Changes in the genetic makeup of the virus may impact its baseline transmissibility. Therefore, for our analysis, we use November 30, 2021 as a cutoff date to assess the impact of the intervention in BC on reducing the spread of SARS-CoV-2. With a population size of 5.07 million in BC \citep{datacommons}, the daily active cases are transformed into weekly infectious proportions, and Bayesian inference was performed based on these observed values. \added{We consider two different identification rates in British Columbia’s COVID-19 data, addressing the variability in testing capacity when the antigen test was approved by Health Canada on October 6, 2022 \citep{CBCAntigenTest}. Let $\tilde{p}_t$ denote the identification rate at time $t$. Two different identification rates are written as
    \begin{equation}
    \tilde{p}_t = \begin{cases}
            p_1, & \text{ if } t \le T^*, \\
            p_2, & \text{ if } t > T^*,
         \end{cases}
         \label{eq: different identification rates}
    \end{equation}
where $T^*$ represents the week of October 6, 2022, marking the approval of the antigen test. } Given the abundance of prior information available for COVID-19 \citep{wangping2020extended,dehning2020inferring, kobayashi2020predicting, CDC2023}, we specify the hyperparameters in the prior distribution as follows:
\begin{equation}
\label{eq: prior for real data analysis}
\begin{aligned}
\alpha & \sim TN(1.4, 0.5^2, 0, +\infty), \\
\beta & \sim TN(\added{1.4}, 0.5^2, 0, +\infty),\\
\gamma & \sim TN(1.4, 0.5^2, 0, +\infty), \\
\lambda & \sim \text{Gamma}(20, 0.01), \\
\kappa & \sim \text{Gamma}(200, 0.01), \\
\text{\added{$p_1$}} & \text{\added{$\sim TN(0.2, 0.05^2, 0.1, 0.4)$}}, \\
\text{\added{$p_2$}} & \text{\added{$\sim TN(0.3, 0.05^2, 0.2, 1)$}}, \\
\end{aligned}
\end{equation}
where the prior variances are relatively larger to fully explore the posterior distribution. The mean of latency parameter $\alpha$ implies an average incubation period of 5 days between the time of exposure and onset of disease symptoms. Similarly, the mean value of the recovery parameter $\gamma$ implies an average period of 5 days between the onset of infectiousness and recovery. The mean values of $\beta$ and $\gamma$ implies a mean value of $R_0$ at \added{1.0} (i.e., \added{1.4}/1.4). Each infectious individual transmits coronavirus to another individual per week on average. Informative Dirichlet priors are used for $\boldsymbol{\pi}_k$ as in the simulation study, such that the transmission rate is more likely to remain at a certain level rather than fluctuating randomly. Metropolis-Hastings within the Particle Gibbs makes it possible to efficiently simulate model parameters from their conditional posterior distributions. Step sizes are adjusted to ensure the acceptance rate is larger than 30\,\%. 

\begin{table}[ht]
\centering
\caption{Comparison of marginal log likelihoods for models with different number of regimes. Standard deviations are given in parenthesis.}
\begin{tabular}{ll}
\hline
\textbf{Model} & $\mathbf{\textbf{log} \ \hat{p}(y_{1:T}| \textbf{Model})}$ \\ 
\hline
$K=1$ & 210.241 (8.473) \\ 
$K=2$ & 490.888 (4.895) \\
$K=3$ & 437.006 (5.462) \\ 
$K=4$ & 424.809 (5.266) \\ 
\hline
\end{tabular}
\label{table: real data analyais - formal model selection}
\end{table}

In Table \ref{table: real data analyais - formal model selection}, we compare the no-switching, two-regime switching, three-regime switching, and four-regime switching models through formal Bayesian model selection. For each setting, we ran 10000 particle MCMC iterations after a burn-in of 1000 iterations. The number of particles for each value of $x_t$ was set to $M=50$. \added{The posterior mean of marginal log-likelihoods from each iteration of CSMC is computed to choose the optimal number of regimes.} We observed a higher data likelihood for the two-regime switching state-space model. Increasing the number of regimes from two to four reduces the model likelihood gradually. We therefore ran two MCMC chains, each consisting of \added{100000} MCMC iterations, after a burn-in period of 1000 iterations. \added{In each iteration, we performed five Metropolis-Hastings updates for the parameters before updating the latent states using CSMC.} This allowed us to efficiently infer parameters and latent variables under a two-regime setting using the weekly observed infections in BC. \added{The convergence of the MCMC chains is thoroughly assessed through multiple diagnostic measures. Trace plots and kernel density plots for each parameter were inspected across the two MCMC chains, as displayed in Figure \ref{fig: two-regime real data analysis trace plot}. Visual examination of these plots revealed stable and well-behaved patterns, indicating convergence. Additionally, the Gelman-Rubin statistic was computed for each parameter. More details are provided in Table \ref{table: GR for two-regime real data analysis}. The values obtained were close to 1, suggesting no non-convergence issues. The combination of visual inspection and formal statistical assessment provides robust evidence that the MCMC chains have successfully converged to the target distribution.}

\begin{figure}[ht]
    \centering
    \includegraphics[scale=0.1]{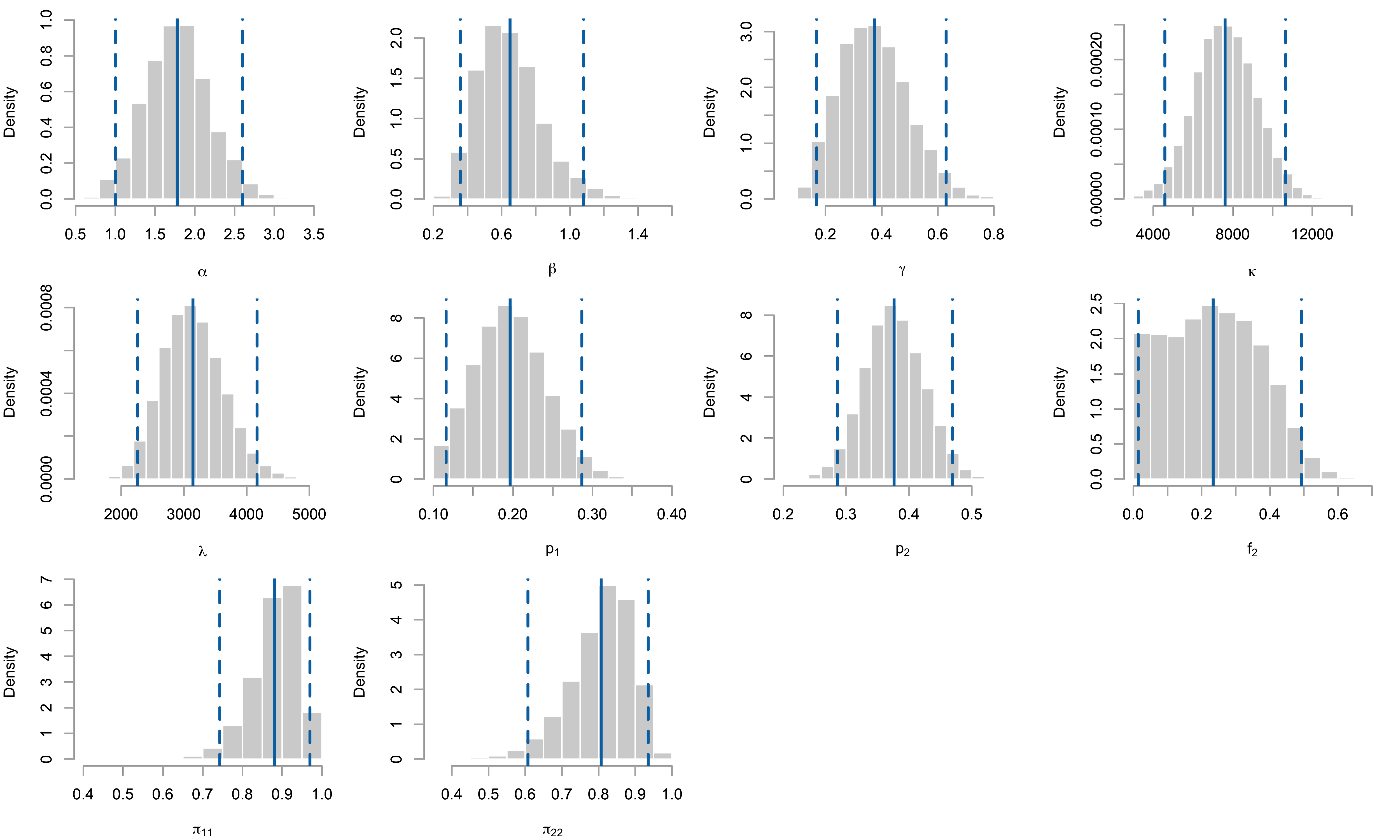}
    \caption{Posterior densities for estimated model parameters in BC weekly data under the two-regime BDSSS-SEIR model. Mean and 95\,\% credible intervals of the posterior densities are shown in solid and dashed blue lines.}
    \label{fig: real data analysis - parameter histogram}
\end{figure}

\begin{figure}[ht]
    \centering
    \includegraphics[scale=0.2]{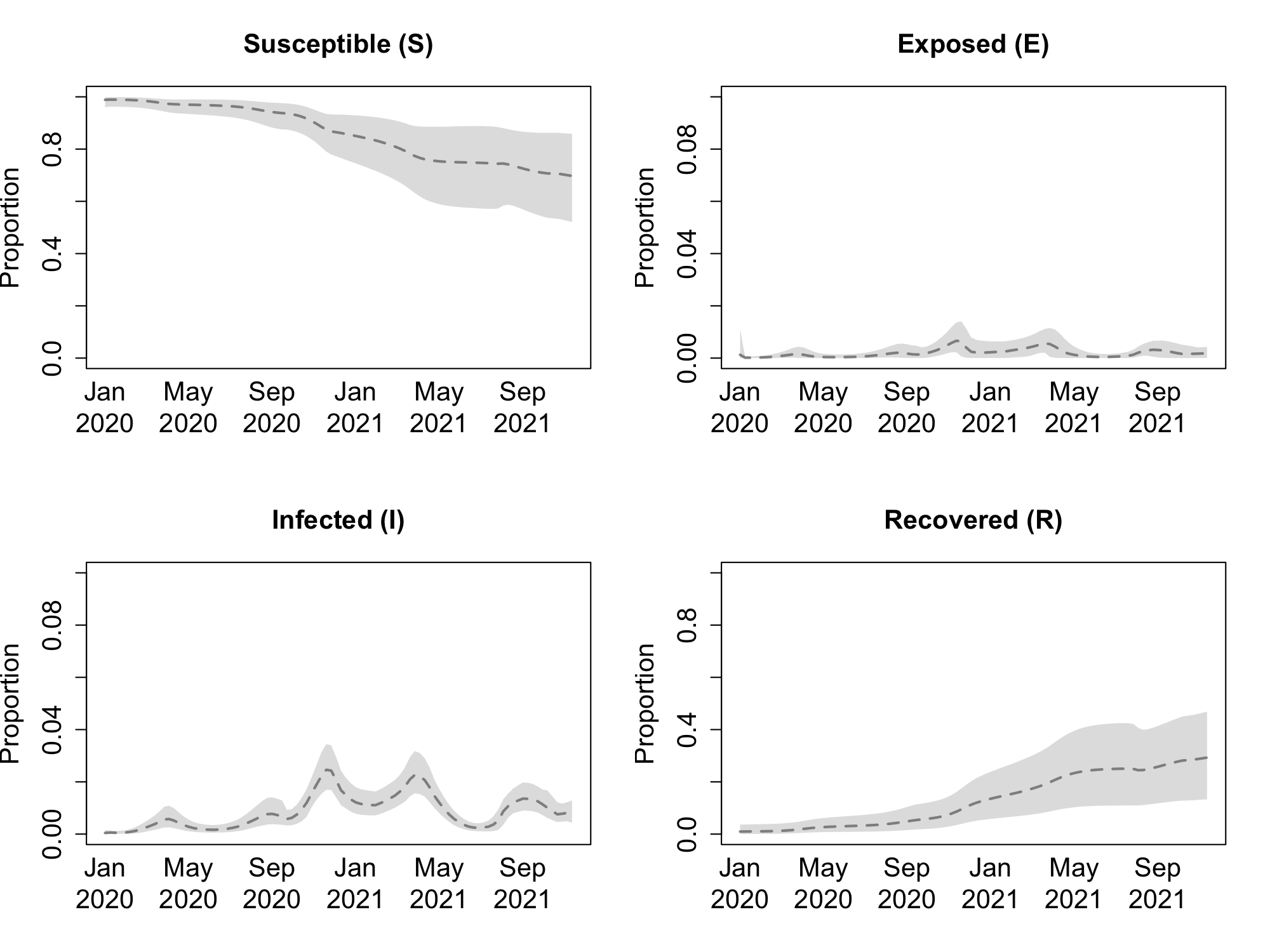}
    \caption{Posterior estimates of SEIR dynamics based on BC COVID-19 weekly active proportions in a two-regime setting. The posterior mean and 95\,\% credible intervals are drawn in dashed grey lines with shaded area.}
    \label{fig: real data analysis - estimated seir}
\end{figure}

Table \ref{table: real data analysis - posterior parameter estimates} reports the posterior mean, median, standard deviation, and 95\,\% credible intervals of the parameter estimates. The posterior distributions of estimated model parameters are provided in Figure \ref{fig: real data analysis - parameter histogram}. For the COVID-19 in BC, the estimated incubation period is \added{0.562 (0.384, 0.997)} weeks, and the recovery period is estimated to be \added{2.674 (1.588, 5.948)} weeks. The baseline transmission rate is estimated to be  \added{0.649 (0.358, 1.081)}. Given the posterior estimates of the transmission rate and the recovery rate, the estimated mean of the basic reproductive number is  \added{1.735}, which lies in the range of 1.4 to 6.49 reported in published studies \citep{liu2020reproductive}. \added{After the approval of antigen test in Canada \citep{CBCAntigenTest}, the estimated identification rate increased from 0.196 (0.116, 0.287) to 0.376 (0.286, 0.470).} The transmission rate modifier in the second regime reduces the transmission rate by \added{76.6\,\% (50.7\,\%, 98.6\,\%)}. Conditional on the posterior summary statistics of $\pi_{11}$ and $\pi_{22}$, the probability of transitioning from one regime to another can be obtained. At time $t$, the probability of transitioning from regime 1 to regime 2 is  \added{0.119 (0.030, 0.257)}, while the probability of transitioning from regime 2 to regime 1 is estimated to be \added{0.193 (0.065, 0.393)}. This suggests that staying in the same regime is more likely than changing abruptly at any time point. The estimated latent SEIR trajectories are shown in Figure \ref{fig: real data analysis - estimated seir}. A gradual decrease is observed in the susceptible proportion, reaching 50\,\% by February 8, 2022, indicating that approximately half of the population in British Columbia has been infected.

\begin{table}[ht]
\centering
\caption{Posterior mean, median, standard deviation (SD), and 95\,\% credible intervals of parameter estimates for B.C COVID-19 data in a two-regime setting.}
\begin{tabular}{lllll}
\hline
\textbf{Parameter} & \textbf{Mean} & \textbf{Median} & \textbf{SD} & \textbf{95\% Credible Interval}\\
\hline
$\alpha$ & 1.779 & 1.773 & 0.403 & (1.003, 2.601) \\
$\beta$ & 0.649 & 0.628  & 0.185 & (0.358, 1.081) \\
$\gamma$ & 0.374 & 0.366 & 0.121 & (0.168, 0.630) \\
$\kappa$ &  7609.821 & 7601.557 & 1546.047 & (4580.536, 10655.747) \\
$\lambda$ & 3144.528 &  3118.411 & 485.455 & (2264.206, 4165.609) \\
$p_1$ & 0.196 & 0.195 & 0.044 & (0.116, 0.287) \\
$p_2$ & 0.376 & 0.375 & 0.047 & (0.286, 0.470) \\
$f_2$ & 0.234 & 0.232 & 0.136 & (0.014, 0.493) \\ 
$\pi_{11}$ & 0.881 & 0.891 & 0.059 & (0.743, 0.970) \\ 
$\pi_{22}$ & 0.807 & 0.820 & 0.085 & (0.607, 0.935) \\ 
\hline
\end{tabular}

\label{table: real data analysis - posterior parameter estimates}
\end{table}

\begin{figure}[ht!]
    \centering
    \includegraphics[scale=0.8]{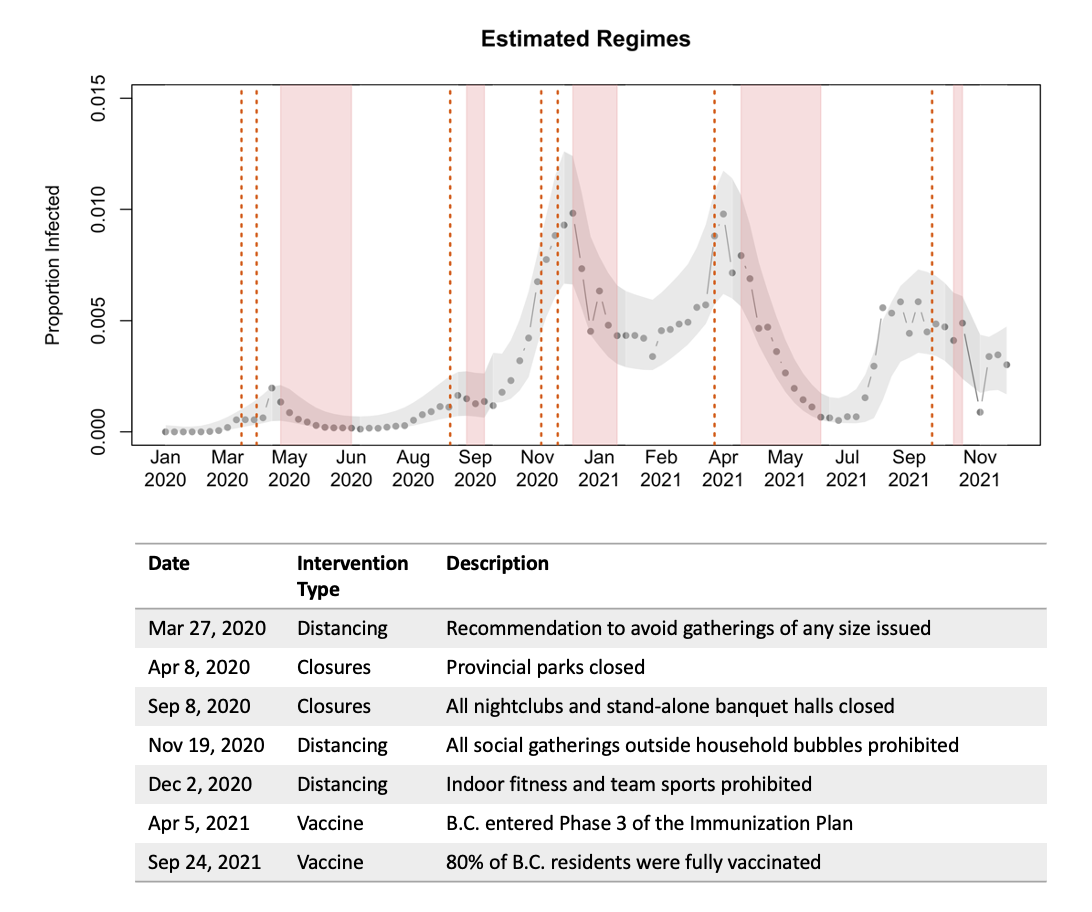}
    \caption{(Top) Posterior estimates of regimes over time for BC weekly data from the week of Jan 27, 2020 to the week of Feb 7, 2022. The grey dots represent the observed $y_t$ every week. The dashed grey curve with shaded area represents the posterior mean and 95\,\% credible interval of the observed infectious proportion. A reduction in the transmission rate is detected during the period with red background, while the transmission rate remains unchanged during other periods. (Bottom) The table presents COVID-19 intervention timeline in BC, as indicated by vertical red lines. }
    \label{fig: real data analysis - estimated Xt}
\end{figure}

Figure \ref{fig: real data analysis - estimated Xt} displays corresponding tracking results of the weekly active proportions from the week of Jan 28, 2020 to the week of Feb 9, 2022 in British Columbia. The graph shows the estimated switching states over time, indicating when a change in the transmission rate occurred. The red background indicates a period where there was a \added{76.6\,\%} reduction in the transmission rate, following external interventions implemented by the BC government to control the spread of COVID-19. The red dashed lines represent the dates when these interventions were implemented. The interventions include the recommendation to avoid gatherings of any size, the closing of provincial parks, \added{closing of nightclubs and stand-alone banquet halls,} prohibition of social gatherings outside household bubbles, prohibition of indoor fitness and team sports, the implementation of the immunization plan to the general public, \added{and the vaccination milestone} \citep{valandos, CIHI}. The results show that the reduction in transmission rate coincides with the implementation of these interventions, providing evidence that these measures were effective in controlling the spread of COVID-19. We also found that it took approximately two to three weeks for the governmental interventions to take effect. The recommendation to avoid gatherings of any size and the closure of provincial parks in March and April 2020 reduced transmission rate for more than two months. Phase 3 of B.C.'s reopening started in June 2020, causing an increase of infected proportions. \added{The closure of nightclubs and stand-alone banquet halls starting on September 8, 2020 also contributed to preventing the spread of COVID-19, although the effect lasted for only three weeks.}  Distancing measures in November and December 2020, including prohibition of social gatherings outside household bubbles, indoor fitness, and team sports, had resulted in a lower transmission rate in the following two months. Phase 3 of BC's COVID-19 immunization plan began in April 2021. Since then, all eligible adults in BC were able to book vaccine appointments online to get their doses. This vaccine intervention effectively reduced the transmission rate in middle 2021. \added{As of September 24, 2021, the number of fully vaccinated residents in British Columbia has reached 80\,\%. Following this vaccination milestone, the transmission rate has decreased.} In summary, these real data analysis results demonstrate the utility of our BDSSS-SEIR model in providing valuable insights into the dynamics of COVID-19 transmission and the effectiveness of external interventions.

\section{Conclusion}
\label{section: Discussion}

We have presented a Beta-Dirichlet switching state-space SEIR model to effectively assess the impact of interventions while tracking the dynamics of an infectious disease. Our model incorporates a discretized ordinary differential equation to capture the underlying SEIR proportions and utilizes a Beta-Dirichlet structure to model the switching mechanism. We employed an advanced Bayesian inference method, particle MCMC, to sample latent variables and explore high-dimensional parameters. The estimation ability of the particle MCMC algorithm on the proposed model has been demonstrated through two-regime and three-regime simulation studies. We also related the inferred change of regimes to the COVID-19 timeline of governmental interventions in British Columbia, Canada. A \added{76.6\,\%} reduction of the transmission rate was detected following the interventions in BC, such as social distancing, venue closures, and implementation of vaccinations, confirming their effectiveness in controlling the spread of COVID-19.

Compared to existing studies about the effectiveness of interventions \citep{wangping2020extended, dehning2020inferring}, a key contribution of our model is the incorporation of a switching state variable that automatically tracks the dynamic change of transmission rate due to external interventions. The switching state variable attached to the transmission rate modifier plays a key role in identifying the timing and magnitude of changes in the transmission rate, making the model more adaptable to changes in the spread of the disease. Furthermore, the utilization of particle MCMC in disease dynamics is an underexplored area in practice \citep{endo2019introduction}. Although MCMC is frequently used in the epidemiological field,  particle MCMC generally outperforms MCMC when handling high-dimensional state spaces and nonlinear or non-Gaussian models \added{\citep{andrieu2010particle, rasmussen2011inference, endo2019introduction}}. In numerical examples, particle MCMC demonstrates efficient inference without requiring a large number of particles or iterations. Moreover, it enables the estimation of the marginal likelihood of the model, facilitating model comparison for determining the optimal number of regimes.

However, implementing the proposed model regularly on real data poses challenges. One challenge is that the prior distribution of parameters needs to be more informative to avoid capturing noises in the data while accounting for a higher number of regimes. As the number of regimes increases, the complexity of parameter estimation grows, which may lead to difficulty in finding the underlying signal if the prior distribution is not informative enough. Second, the model assumes that the transmission rate changes solely due to interventions and may not account for other factors influencing the spread of the disease, such as changes in human behavior or the emergence of new variants of the virus. For example, a new variant of the virus that is more transmissible could increase the transmission rate abruptly, even if interventions remain unchanged. Since the transmission rate in our model is typically assumed to account for a single type of virus, we require users to consider distinct baseline transmission rates for different strains of the virus. Therefore, these extra factors should also be considered when making informed public health decisions.

\appendix

\clearpage
\section{Appendix}

\subsection{Derivation of importance weights}
\label{sec: Derivation of Importance Weights}
The particle weights play a crucial role for weighting the particles in SMC. Particles with higher weights are more likely to be resampled and propagated forward to the next time step. In general, the un-normalized importance weights are calculated as the ratio of the target density over the proposal density
\begin{align} 
\label{eq: un-normalized importance weight}
   w_t^{(i)} = \frac{p_\psi(\boldsymbol{\theta}_{1:t}^{(i)}, x_{1:t}^{(i)} | y_{1:t})}{q_\psi(\boldsymbol{\theta}_{1:t}^{(i)}, x_{1:t}^{(i)} | y_{1:t})}, \ \ i=1,...,N. 
\end{align}
The basic idea behind importance sampling is to use the proposal distribution to generate particles that can be used to estimate statistical properties of the target distribution. However, since the particles are drawn from a different distribution, they need to be weighted in order to account for the difference between the proposal and the target distribution. The importance weight $w_t^{(i)}$ indicates how much the $i$-th particle set contributes to the estimation of the target distribution at time $t$. The higher the importance weight, the more important the particle is in estimating the target distribution. In practice, a proposal distribution similar to target is preferred so that a finite number of weighted particles estimate the target distribution closely. By using Bayes theorem and Markov property, the target density can be factorized as
\begin{equation}
\label{eq:numerator}
\begin{aligned}
p_{\psi}\left(\boldsymbol{\theta}_{1: t}, x_{1: t} \mid y_{1: t}\right)
&=p_{\psi}\left(\boldsymbol{\theta}_{t}, \boldsymbol{\theta}_{1: t-1}, x_{t}, x_{1: t-1} \mid y_{t}, y_{1: t-1}\right) \\
&=\frac{p_{\psi}\left(y_{t}, \boldsymbol{\theta}_{t}, \boldsymbol{\theta}_{1: t-1}, x_{t}, x_{1: t-1} \mid y_{1: t-1}\right)}{p_{\psi}\left(y_{t} \mid y_{1: t-1}\right)} \\
&=\frac{h_{\psi}\left(y_{t} \mid \boldsymbol{\theta}_{1: t}, x_{1: t}\right) g_{\psi}\left(\boldsymbol{\theta}_{t} \mid \boldsymbol{\theta}_{1: t-1}, x_{1: t}\right) p_{\psi}\left(x_{t} \mid x_{t-1}\right)}{p_{\psi}\left(y_{t} \mid y_{1: t-1}\right)} p_{\psi}\left(\boldsymbol{\theta}_{1: t-1}, x_{1: t-1} \mid y_{1: t-1}\right) \\
&=\frac{h_{\psi}\left(y_{t} \mid \boldsymbol{\theta}_{t}, x_{ t}\right) g_{\psi}\left(\boldsymbol{\theta}_{t} \mid \boldsymbol{\theta}_{t-1}, x_{t}\right) p_{\psi}\left(x_{t} \mid x_{t-1}\right)}{p_{\psi}\left(y_{t} \mid y_{1: t-1}\right)} p_{\psi}\left(\boldsymbol{\theta}_{1: t-1}, x_{1: t-1} \mid y_{1: t-1}\right) \\
& \propto h_{\psi}\left(y_{t} \mid \boldsymbol{\theta}_{t}, x_{ t}\right) g_{\psi}\left(\boldsymbol{\theta}_{t} \mid \boldsymbol{\theta}_{t-1}, x_{t}\right) p_{\psi}\left(x_{t} \mid x_{t-1}\right) p_{\psi}\left(\boldsymbol{\theta}_{1: t-1}, x_{1: t-1} \mid y_{1: t-1}\right),
\end{aligned}
\end{equation}
and the proposal density can be written as
\begin{equation}
\label{eq:denominator}
    \begin{aligned}
        q_\psi \left(\boldsymbol{\theta}_{1: t}, x_{1:t} \mid y_{1: t}\right) & = q_\psi \left(\boldsymbol{\theta}_t, x_t \mid \boldsymbol{\theta}_{1: t-1}, x_{1:t-1}, y_{1: t}\right) q_\psi \left(\boldsymbol{\theta}_{1: t-1}, x_{1:t-1} \mid y_{1: t}\right) \\
        & \propto q_\psi \left(\boldsymbol{\theta}_t, x_t \mid \boldsymbol{\theta}_{t-1}, x_{t-1}, y_t\right) q_\psi \left(\boldsymbol{\theta}_{1: t-1}, x_{1:t-1} \mid y_{1: t-1}\right),
    \end{aligned}
\end{equation}
where the indices of particle are ignored for brevity. Note that $q_\psi \left(\boldsymbol{\theta}_{1: t}, x_{1:t} \mid y_{1: t}\right)$ is defined on the same domain as the target distribution $p_\psi \left(\boldsymbol{\theta}_{1: t}, x_{1: t}\mid y_{1: t}\right)$, but generally $q_\psi (.)$ is not required to satisfy Markovian property because it might be a density defined outside the switching state-space model. Plugging (\ref{eq:numerator}) and (\ref{eq:denominator}) into un-normalized importance weight in (\ref{eq: un-normalized importance weight}), we obtain
\begin{equation}
    \begin{aligned}
    w_t^{(i)} & \propto \frac{p_\psi \left(\boldsymbol{\theta}_{1: t-1}^{(i)}, x_{1:t-1}^{(i)} \mid y_{1: t-1}\right)}{q_\psi \left(\boldsymbol{\theta}_{1: t-1}^{(i)}, x_{1:t-1}^{(i)} \mid y_{1: t-1}\right) } \times \frac{h_{\psi} (y_{t} \mid \boldsymbol{\theta}_{t}^{(i)}, x_{t}^{(i)}) g_{\psi}\left(\boldsymbol{\theta}_{t}^{(i)} \mid \boldsymbol{\theta}_{t-1}^{(i)}, x_{t}^{(i)}\right) p_{\psi}\left(x_{t}^{(i)} \mid x_{t-1}^{(i)}\right)}{q_\psi(\boldsymbol{\theta}_{t}^{(i)}, x_{t}^{(i)} \mid \boldsymbol{\theta}_{t-1}^{(i)}, x_{t-1}^{(i)}, y_{t})} \\
    & =  w_{t-1}^{(i)} \times \frac{h_{\psi} (y_{t} \mid \boldsymbol{\theta}_{t}^{(i)}, x_{t}^{(i)}) g_{\psi}\left(\boldsymbol{\theta}_{t}^{(i)} \mid \boldsymbol{\theta}_{t-1}^{(i)}, x_{t}^{(i)}\right) p_{\psi}\left(x_{t}^{(i)} \mid x_{t-1}^{(i)}\right)}{q_\psi(\boldsymbol{\theta}_{t}^{(i)}, x_{t}^{(i)} \mid \boldsymbol{\theta}_{t-1}^{(i)}, x_{t-1}^{(i)}, y_{t})} \\
    & = w_{t-1}^{(i)} \bar{w}_t^{(i)},
    \end{aligned}
\end{equation}
where $\bar{w}_t^{(i)}$ is the so-called incremental importance weight. In SMC, the incremental importance weights plays a crucial role because of guiding the selection of particles for resampling and propagation to the next time step, and ensuring that the particle weights accurately represent the posterior distribution over time. Once $w_t^{(i)}$ are available, the normalized importance weight $W_t^{(i)}$ can be obtained as
\begin{align}
   W_t^{(i)} = \frac{w_t^{(i)}}{\sum_{j=1}^N w_t^{(j)}}, \ \ i=1,...,N
\end{align}
such that $\sum_{i=1}^N W_t^{(i)} = 1$. At each iteration of SMC, we used the multinomial resampling procedure where particles at time $t$ choose their ancestral particles at time $t-1$ from $\text{Multinomial}(\{W_{t-1}^{(i)}\}_{i=1}^N)$. This is done by simulating the ancestor indices $\{a_t^{(i)}\}_{i=1}^N$ that represents the index of the ancestor of particle $\{\boldsymbol{\theta}_{1:t}^{(i)}, x_{1:t}^{(i)}\}$ at time $t-1$. After resampling at every time step, the equal weights $W_{t-1}^{(i)}=1/N$ and $w_{t-1}^{(i)}=1$ are assigned to resampled particles at time $t-1$. Indeed, we can safely ignore $w_{t-1}^{(i)}$ while computing the normalized weight at time $t$, so that the un-normalized importance weights after resampling step become
\begin{equation} 
\label{eq:un-normalized weight}
    w_t^{(i)} = w_{t-1}^{(i)} \bar{w}_t^{(i)} =  \bar{w}_t^{(i)} = \frac{h_{\psi} (y_{t} \mid \boldsymbol{\theta}_{t}^{(i)}, x_{t}^{(i)}) g_{\psi}\left(\boldsymbol{\theta}_{t}^{(i)} \mid \boldsymbol{\theta}_{t-1}^{(i)}, x_{t}^{(i)}\right) p_{\psi}\left(x_{t}^{(i)} \mid x_{t-1}^{(i)}\right)}{q_\psi(\boldsymbol{\theta}_{t}^{(i)}, x_{t}^{(i)} \mid \boldsymbol{\theta}_{t-1}^{(i)}, x_{t-1}^{(i)}, y_{t})},
\end{equation}
which relies on incremental importance weight only.  In our case, we propogate particles by simulating directly from the state transition density in (\ref{eq: BDSSS-SEIR-Transition Equation}) and observation density in (\ref{eq: BDSSS-SEIR-Observation Equation}) because there is no ideal importance density. With the proposal density 
\begin{equation}
\label{eq: proposal density}
    q_\psi(\boldsymbol{\theta}_{t}^{(i)}, x_{t}^{(i)} \mid \boldsymbol{\theta}_{t-1}^{(i)}, x_{t-1}^{(i)}, y_{t}) = g_{\psi}\left(\boldsymbol{\theta}_{t}^{(i)} \mid \boldsymbol{\theta}_{t-1}^{(i)}, x_{t}^{(i)}\right) p_{\psi}\left(x_{t}^{(i)} \mid x_{t-1}^{(i)}\right),
\end{equation}
the un-normalized importance weights in (\ref{eq:un-normalized weight}) simplify to
\begin{equation}
    w_t^{(i)} \propto h_\psi(y_t|\boldsymbol{\theta}_t^{(i)}, x_t^{(i)})
\end{equation}
for $t=2, ..., T$.

\subsection{Posterior distribution}
\label{sec: posterior distribution}
By incorporating the BDSSS-SEIR model structure, the joint posterior distribution of latent variables and model parameters is written as
\begin{equation}
    \begin{aligned}
    p(\psi,  \boldsymbol{\theta}_{1:T}, & x_{1:T} | y_{1:T}) \propto p(\boldsymbol{\theta}_{1:T}, x_{1:T}, y_{1:T}|\psi) \pi(\psi) \\
    = \bigg[ \prod_{t=1}^T & f_{\psi}(y_t|\boldsymbol{\theta}_{1:t}, x_{1:t}) \prod_{t=2}^T g_{\psi}(\boldsymbol{\theta}_t|\boldsymbol{\theta}_{1:t-1}, x_{1:t}) \prod_{t=2}^T p_{\psi}(x_t|x_{t-1}) \bigg] \mu_\psi(\boldsymbol{\theta}_1 |  x_1) \mu_\psi(x_1)  \\
    & \times \pi(\alpha) \pi(\beta) \pi(\gamma) \pi(\lambda) \pi(\kappa) \pi(p) \pi(P_X) \pi(f_2) \pi(f_3) ... \pi(f_K) \\
    \propto \bigg[ \prod_{t=1}^T & \frac{1}{B(\lambda p I_t, \lambda (1- p I_t))} y_t^{\lambda p I_t-1} (1-y_t)^{\lambda (1- p I_t)-1} 
    \\ & \times \prod_{t=2}^T \frac{1}{B(\kappa \eta^S_t, \kappa \eta^E_t, \kappa \eta^I_t, \kappa \eta^R_t)} S_t^{\kappa \eta^S_t - 1} E_t^{\kappa \eta^E_t-1} I_t^{\kappa \eta^I_t-1} R_t^{\kappa \eta^R_t - 1} \\
    & \times  \prod_{t=2}^T \pi_{ij}^{(t)}  \bigg] \times  \frac{1}{B(100,1,1,1)} S_1^{99}  \times \frac{1}{K} \\
    & \times \frac{1}{\sigma_{\alpha}} \exp{(-\frac{(\alpha-m_{\alpha})^2}{2 \sigma_\alpha^2})} I_{\alpha>0} \times \frac{1}{\sigma_{\beta}} \exp{(-\frac{(\beta - m_{\beta})^2}{2 \sigma_\beta^2})} I_{\beta>0}  \times \frac{1}{\sigma_{\gamma}} \exp{(-\frac{(\gamma-m_{\gamma})^2}{2 \sigma_\gamma^2})} I_{\gamma>0} \\ & \times \lambda^{a_\lambda-1} \exp{(-b_\lambda \lambda)} \times \kappa^{a_\kappa-1} \exp{(-b_\kappa \kappa)} \\ & \times
\prod_{k=1}^K \bigg[ \frac{\Gamma\left(\sum_{d=1}^K \delta_{k d}\right)}{\prod_{d=1}^K \Gamma\left(\delta_{k d}\right)} \prod_{d=1}^K \pi_{k d}^{\delta_{k d}-1} \bigg] \\
& \times (K-1)^{K-1} ,
\end{aligned} 
\end{equation}
where $B$ represents Beta functions
\begin{align*}
    B(\lambda p I_t, \lambda (1- p I_t)) & = \frac{\Gamma (\lambda p I_t) \Gamma(\lambda (1- p I_t))}{\Gamma(\lambda p I_t + \lambda (1- p I_t))} = \frac{\Gamma (\lambda p I_t) \Gamma(\lambda (1- p I_t))}{\Gamma(\lambda)}, \\
    B(\kappa \eta^S_t, \kappa \eta^E_t, \kappa \eta^I_t, \kappa \eta^R_t) & = \frac{\Gamma(\kappa \eta^S_t) \Gamma(\kappa \eta^E_t) \Gamma(\kappa \eta^I_t) \Gamma(\kappa \eta^R_t)}{\Gamma(\kappa \eta^S_t + \kappa \eta^E_t + \kappa \eta^I_t + \kappa \eta^R_t)}, \\
    B(100, 1, 1, 1) &= \frac{\Gamma(100) \Gamma(1) \Gamma(1) \Gamma(1)}{\Gamma(103)}.
\end{align*}

\clearpage
\newpage
\bibliography{main.bib}

\clearpage
\begin{center}
\textbf{\Large Supplementary Material}
\end{center}
\setcounter{equation}{0}
\setcounter{figure}{0}
\setcounter{table}{0}
\setcounter{page}{1}
\setcounter{section}{0}
\makeatletter
\renewcommand{\theequation}{S\arabic{equation}}
\renewcommand{\thefigure}{S\arabic{figure}}
\renewcommand{\thealgorithm}{S\arabic{algorithm}}
\renewcommand{\thesection}{S\arabic{section}}
\renewcommand{\thetable}{S\arabic{table}}
\renewcommand{\bibnumfmt}[1]{[S#1]}
\renewcommand{\citenumfont}[1]{S#1}


\section{The 4-th order Runge-Kutta approximation method}
\label{sec: RK4 details}

The function $r(\boldsymbol{\theta}_{t-1}; \alpha, \beta, \gamma, f_{x_t})$ provides a prescription to propagate the new SEIR model forward in one time unit. It is approximated according to the 4-th order Runge-Kutta (RK4) method \citep{kutta1901beitrag}, as illustrated below:
\begin{equation}
\label{eq:RK4}
r(\boldsymbol{\theta}_{t-1}; \alpha, \beta, \gamma, f_{x_t}) = 
\begin{bmatrix}
\eta^S_t \\
\eta^E_t \\
\eta^I_t \\
\eta^R_t
\end{bmatrix} = 
\begin{bmatrix}
S_{t-1} + 1/6(k_{t-1}^{S_1} + 2k_{t-1}^{S_2} + 2k_{t-1}^{S_3} + k_{t-1}^{S_4}) \\
E_{t-1} + 1/6(k_{t-1}^{E_1} + 2k_{t-1}^{E_2} + 2k_{t-1}^{E_3} + k_{t-1}^{E_4}) \\
I_{t-1} + 1/6(k_{t-1}^{I_1} + 2k_{t-1}^{I_2} + 2k_{t-1}^{I_3} + k_{t-1}^{I_4}) \\
R_{t-1} + 1/6(k_{t-1}^{R_1} + 2k_{t-1}^{R_2} + 2k_{t-1}^{R_3} + k_{t-1}^{R_4})
\end{bmatrix},
\end{equation}
where \\
\begin{align*}
    k_{t-1}^{S_1} & = -f_{x_t} \beta S_{t-1}I_{t-1} \\
    k_{t-1}^{S_2} & = -f_{x_t} \beta(S_{t-1} + 0.5k_{t-1}^{S_1})(I_{t-1} + 0.5k_{t-1}^{I_1}) \\
    k_{t-1}^{S_3} & = -f_{x_t} \beta(S_{t-1} + 0.5k_{t-1}^{S_2})(I_{t-1} + 0.5k_{t-1}^{I_2}) \\
    k_{t-1}^{S_4} & = -f_{x_t} \beta(S_{t-1} + k_{t-1}^{S_3})(I_{t-1} + k_{t-1}^{I_3}) \added{,} \\  \\
    k_{t-1}^{E_1} & = f_{x_t} \beta S_{t-1}I_{t-1} - \alpha E_{t-1}\\
    k_{t-1}^{E_2} & = f_{x_t} \beta(S_{t-1} + 0.5k_{t-1}^{S_1})(I_{t-1} + 0.5k_{t-1}^{I_1}) - \alpha(E_{t-1} + 0.5k_{t-1}^{E_1})\\
    k_{t-1}^{E_3} & = f_{x_t} \beta(S_{t-1} + 0.5k_{t-1}^{S_2})(I_{t-1} + 0.5k_{t-1}^{I_2}) - \alpha(E_{t-1} + 0.5k_{t-1}^{E_2}) \\
    k_{t-1}^{E_4} & = f_{x_t} \beta(S_{t-1} + k_{t-1}^{S_3})(I_{t-1} + k_{t-1}^{I_3}) - \alpha(E_{t-1} + k_{t-1}^{E_3})\added{,} \\ \\
    k_{t-1}^{I_1} & = \alpha E_{t-1} - \gamma I_{t-1}\\
    k_{t-1}^{I_2} & = \alpha(E_{t-1} + 0.5k_{t-1}^{E_1}) - \gamma (I_{t-1} + 0.5k_{t-1}^{I_1})\\
    k_{t-1}^{I_3} & = \alpha(E_{t-1} + 0.5k_{t-1}^{E_2}) - \gamma (I_{t-1} + 0.5k_{t-1}^{I_2})\\
    k_{t-1}^{I_4} & = \alpha(E_{t-1} + k_{t-1}^{E_3}) - \gamma (I_{t-1} + k_{t-1}^{I_3}) \added{,} \\ \\
    k_{t-1}^{R_1} & = \gamma I_{t-1}\\
    k_{t-1}^{R_2} & = \gamma (I_{t-1} + 0.5k_{t-1}^{I_1})\\
    k_{t-1}^{R_3} & = \gamma (I_{t-1} + 0.5k_{t-1}^{I_2})\\
    k_{t-1}^{R_4} & = \gamma (I_{t-1} + k_{t-1}^{I_3}). \\ 
\end{align*}

\clearpage
\section{Metropolis-Hastings algorithm for sampling model parameters} 
\label{sec: MH steps}

\begin{algorithm}[ht]
 \caption{Metropolis-Hastings algorithm for sampling $\alpha$}
 \label{algorithm: MH step for sampling alpha}
\begin{algorithmic}[1]
 \State  Given the current value $\alpha^{(r)}$, simulate $\alpha^* \sim Q(\alpha^*|\alpha^{(r)}) = \text{TN}(\alpha^{(r)}, \sigma_{\alpha^*}^2, 0, \infty)$\;
 \State Compute the acceptance ratio:
  $$A_{\alpha^*|\alpha^{(r)}} = \min\left\{\frac{Q(\alpha^{(r)}|\alpha^*)p_{\alpha^*}(\boldsymbol{\theta}_{1:T}, x_{1:T}, \psi|y_{1:T})}{Q(\alpha^*|\alpha^{(r)})p_{\alpha^{(r)}}(\boldsymbol{\theta}_{1:T}, x_{1:T}, \psi|y_{1:T})}, 1\right\}.$$
  \State Sample $u \sim \text{Unif}(0,1)$.
  Set $ \alpha^{(r+1)} = \begin{cases}
      \alpha^*, & \text{if $u < A_{\alpha^*|\alpha^{(r)}}$},\\
      \alpha^{(r)}, & \text{otherwise}.
    \end{cases} $
\end{algorithmic}
\end{algorithm}

\begin{algorithm}[ht]
 \caption{Metropolis-Hastings algorithm for sampling $\beta$}
 \label{algorithm: MH step for sampling beta}
\begin{algorithmic}[1]
 \State Given the current value $\beta^{(r)}$, simulate $\beta^* \sim Q(\beta^{*}|\beta^{(r)}) = \text{TN}(\beta^{(r)}, \sigma_{\beta^*}^2, 0, \infty)$ 
 \State Compute the acceptance ratio
        $$ A_{\beta^*|\beta^{(r)}} = \text{min}\left\{\frac{ Q(\beta^{(r)}|\beta^{*})p_{\beta^{*}}(\boldsymbol{\theta}_{1:T}, x_{1:T}, \psi|y_{1:T}) }{ Q(\beta^{*}|\beta^{(r)}) p_{\beta^{(r)}}(\boldsymbol{\theta}_{1:T}, x_{1:T}, \psi|y_{1:T}) }, 1\right\} $$
 \State Sample $u \sim \text{Unif}(0,1)$. Set
         $\beta^{(r+1)}=\begin{cases}
         \beta^*, & \text{if $u<A_{\beta^*|\beta^{(r)}}$}, \\
         \beta^{(r)}, & \text{otherwise}.
         \end{cases} $
\end{algorithmic}
\end{algorithm}

\begin{algorithm}[H]
 \caption{Metropolis-Hastings algorithm for sampling $\gamma$}
 \label{algorithm: MH step for sampling gamma}
\begin{algorithmic}[1]
 \State Given the current value $\gamma^{(r)}$, simulate $\gamma^* \sim Q(\gamma^{*}|\gamma^{(r)}) = \text{TN}(\gamma^{(r)}, \sigma_{\gamma^*}^2, 0, \infty)$.
 \State Compute the acceptance ratio
        $$
        A_{\gamma^*|\gamma^{(r)}} = \text{min}\left\{\frac{ Q(\gamma^{(r)}|\gamma^{*})p_{\gamma^{*}}(\boldsymbol{\theta}_{1:T}, x_{1:T}, \psi|y_{1:T}) }{ Q(\gamma^{*}|\gamma^{(r)}) p_{\gamma^{(r)}}(\boldsymbol{\theta}_{1:T}, x_{1:T}, \psi|y_{1:T}) }, 1\right\} 
        $$
  \State Sample $u \sim \text{Unif}(0,1)$. Set
         $ \gamma^{(r+1)} = \begin{cases}
         \gamma^*, & \text{if $u<A_{\gamma^*|\gamma^{(r)}}$}.\\
         \gamma^{(r)}, & \text{otherwise}.
         \end{cases}$
\end{algorithmic}
\end{algorithm}

\begin{algorithm}[H]
 \caption{Metropolis-Hastings algorithm for sampling $\lambda$}
 \label{algorithm: MH step for sampling lambda}
\begin{algorithmic}[1]
 \State Given the current value $\lambda^{(r)}$, simulate $\lambda^* \sim Q(\lambda^{*}|\lambda^{(r)}) = \text{TN}(\lambda^{(r)}, \sigma_{\lambda^*}^2, 0, \infty)$.
 \State Compute the acceptance ratio
        $$
        A_{\lambda^*|\lambda^{(r)}} = \text{min}\left\{\frac{ Q(\lambda^{(r)}|\lambda^{*})p_{\lambda^{*}}(\boldsymbol{\theta}_{1:T}, x_{1:T}, \psi|y_{1:T}) }{ Q(\lambda^{*}|\lambda^{(r)}) p_{\lambda^{(r)}}(\boldsymbol{\theta}_{1:T}, x_{1:T}, \psi|y_{1:T}) }, 1\right\} 
        $$
  \State Sample $u \sim \text{Unif}(0,1)$. Set
         $ \lambda^{(r+1)} = \begin{cases}
         \lambda^*, & \text{if $u<A_{\lambda^*|\lambda^{(r)}}$}.\\
         \lambda^{(r)}, & \text{otherwise}.
         \end{cases}$
\end{algorithmic}
\end{algorithm}

\begin{algorithm}[H]
 \caption{Metropolis-Hastings algorithm for sampling $\kappa$}
 \label{algorithm: MH step for sampling kappa}
\begin{algorithmic}[1]
 \State Given the current value $\kappa^{(r)}$, simulate $\kappa^* \sim Q(\kappa^{*}|\kappa^{(r)}) = \text{TN}(\kappa^{(r)}, \sigma_{\kappa^*}^2, 0, \infty)$.
 \State Compute the acceptance ratio
        $$
        A_{\kappa^*|\kappa^{(r)}} = \text{min}\{\frac{ Q(\kappa^{(r)}|\kappa^{*})p_{\kappa^{*}}(\boldsymbol{\theta}_{1:T}, x_{1:T}, \psi|y_{1:T}) }{ Q(\kappa^{*}|\kappa^{(r)}) p_{\kappa^{(r)}}(\boldsymbol{\theta}_{1:T}, x_{1:T}, \psi|y_{1:T}) }, 1\} 
        $$
  \State Sample $u \sim \text{Unif}(0,1)$. Set
         $ \kappa^{(r+1)} = \begin{cases}
         \kappa^*, & \text{if $u<A_{\kappa^*|\kappa^{(r)}}$}.\\
         \kappa^{(r)}, & \text{otherwise}.
         \end{cases}$
\end{algorithmic}
\end{algorithm}

\begin{algorithm}[H]
 \caption{Metropolis-Hastings algorithm for sampling $p$}
 \label{algorithm: MH step for sampling p}
\begin{algorithmic}[1]
 \State Given the current value $p^{(r)}$, simulate $p^* \sim Q(p^{*}|p^{(r)}) = \text{TN}(p^{(r)}, \sigma_{p^*}^2, 0, \infty)$.
 \State Compute the acceptance ratio
        $$
        A_{p^*|p^{(r)}} = \text{min}\{\frac{ Q(p^{(r)}|p^{*})p_{p^{*}}(\boldsymbol{\theta}_{1:T}, x_{1:T}, \psi|y_{1:T}) }{ Q(p^{*}|p^{(r)}) p_{p^{(r)}}(\boldsymbol{\theta}_{1:T}, x_{1:T}, \psi|y_{1:T}) }, 1\} 
        $$
  \State Sample $u \sim \text{Unif}(0,1)$. Set
         $ p^{(r+1)} = \begin{cases}
         p^*, & \text{if $u<A_{p^*|p^{(r)}}$}.\\
         p^{(r)}, & \text{otherwise}.
         \end{cases}$
\end{algorithmic}
\end{algorithm}

\begin{algorithm}[H]
 \caption{Metropolis-Hastings algorithm for sampling $f_{x_t}, \text{where } x_t=2,3,\ldots,K$}
 \label{algorithm: MH step for sampling f_{x_t}}
\begin{algorithmic}[1]
 \State Given the current value $f_{x_t}^{(r)}$, simulate $f_{x_t}^* \sim Q(f_{x_t}^{*}|f_{x_t}^{(r)}) = \text{TN}(f_{x_t}^{(r)}, \sigma_{f_{x_t}^*}^2, a_{f_{x_t}}, b_{f_{x_t}})$, where $a_{f_{x_t}}$ and $b_{f_{x_t}}$ correspond to the boundary values in the Uniform prior distribution of $f_{x_t}$.
 \State Compute the acceptance ratio
        $$
        A_{f_{x_t}^*|f_{x_t}^{(r)}} = \text{min}\left\{\frac{ Q(f_{x_t}^{(r)}|f_{x_t}^{*})p_{f_{x_t}^{*}}(\boldsymbol{\theta}_{1:T}, x_{1:T}, \psi|y_{1:T}) }{ Q(f_{x_t}^{*}|f_{x_t}^{(r)}) p_{f_{x_t}^{(r)}}(\boldsymbol{\theta}_{1:T}, x_{1:T}, \psi|y_{1:T}) }, 1\right\} 
        $$
  \State Sample $u \sim \text{Unif}(0,1)$. Set
         $ f_{x_t}^{(r+1)} = \begin{cases}
         f_{x_t}^*, & \text{if $u<A_{f_{x_t}^*|f_{x_t}^{(r)}}$}.\\
         f_{x_t}^{(r)}, & \text{otherwise}.
         \end{cases}$
\end{algorithmic}
\end{algorithm}

\begin{algorithm}[H]
 \caption{Metropolis-Hastings algorithm for sampling $\boldsymbol{P}_X=[\boldsymbol{\pi}_{1} \hdots \boldsymbol{\pi}_{K}]^\top$}
 \label{algorithm: MH step for sampling P_X}
\begin{algorithmic}[1]
\State Sample $u' \sim \text{Unif}(0,1)$. Set $k = \lceil u'(K) \rceil$.
\State Update $\boldsymbol{\pi}_k = [\pi_{k1}, \ldots, \pi_{kK}]$ as follows: 
\begin{itemize}
    \item Given the current value $\boldsymbol{\pi}_{k}^{(r)}= [\pi_{k1}^{(r)}, \ldots, \pi_{kK}^{(r)}]$, draw independent samples $[\pi_{k1}^*, \ldots, \pi_{k(K-1)}^*]$
    \begin{align*}
    \pi_{k1}^* \sim Q(\pi_{k1}^*|\pi_{k1}^{(r)}) & = \text{TN}(\pi_{k1}^{(r)}, \sigma^2_{\pi_{k1}^*}, 0, 1) \\
    & \vdots \\
    \pi_{k(K-1)}^* \sim Q(\pi_{k(K-1)}^*|\pi_{k(K-1)}^{(r)}) &= \text{TN}(\pi_{k(K-1)}^{(r)}, \sigma^2_{\pi_{k(K-1)}^*}, 0, 1- \pi_{k(K-2)}^*) \text{ for } K \ge 3
    \end{align*}
    and compute $\pi_{kK}^* = 1- \sum_{j=1}^{K-1} \pi_{kj}^*$ deterministically. Set new parameter $\boldsymbol{\pi}_{k}^{*}= [\pi_{k1}^*, \ldots, \pi_{kK}^*].$
    \item Compute the acceptance ratio 
    $$r = \text{min} \left\{\frac{Q(\pi_{k1}^{(r)}|\pi_{k1}^*) \hdots Q(\pi_{k(K-1)}^{(r)}|\pi_{k(K-1)}^*)p_{\boldsymbol{\pi}_k^{*}}(\boldsymbol{\theta}_{1:T}, x_{1:T}, \psi|y_{1:T}))}{ Q(\pi_{k1}^*|\pi_{k1}^{(r)}) \hdots Q(\pi_{k(K-1)}^*|\pi_{k(K-1)}^{(r)}) p_{\boldsymbol{\pi}_k^{(r)}}(\boldsymbol{\theta}_{1:T}, x_{1:T}, \psi|y_{1:T})} , 1 \right\}$$
    \item Sample $u \sim \text{Unif}(0,1)$. Set 
    $\boldsymbol{\pi}_{k}^{(r+1)}=\begin{cases}
    \boldsymbol{\pi}_{k}^*, & \text{if $u<r$}.\\
    \boldsymbol{\pi}_{k}^{(r)}, & \text{otherwise}.
  \end{cases}$
\end{itemize}
\end{algorithmic}
\end{algorithm}

\clearpage
\section{Trace plots and kernel density plots}
\label{sec: trace plot}

\begin{figure}[ht!]
\centering
\includegraphics[scale=0.088]{ 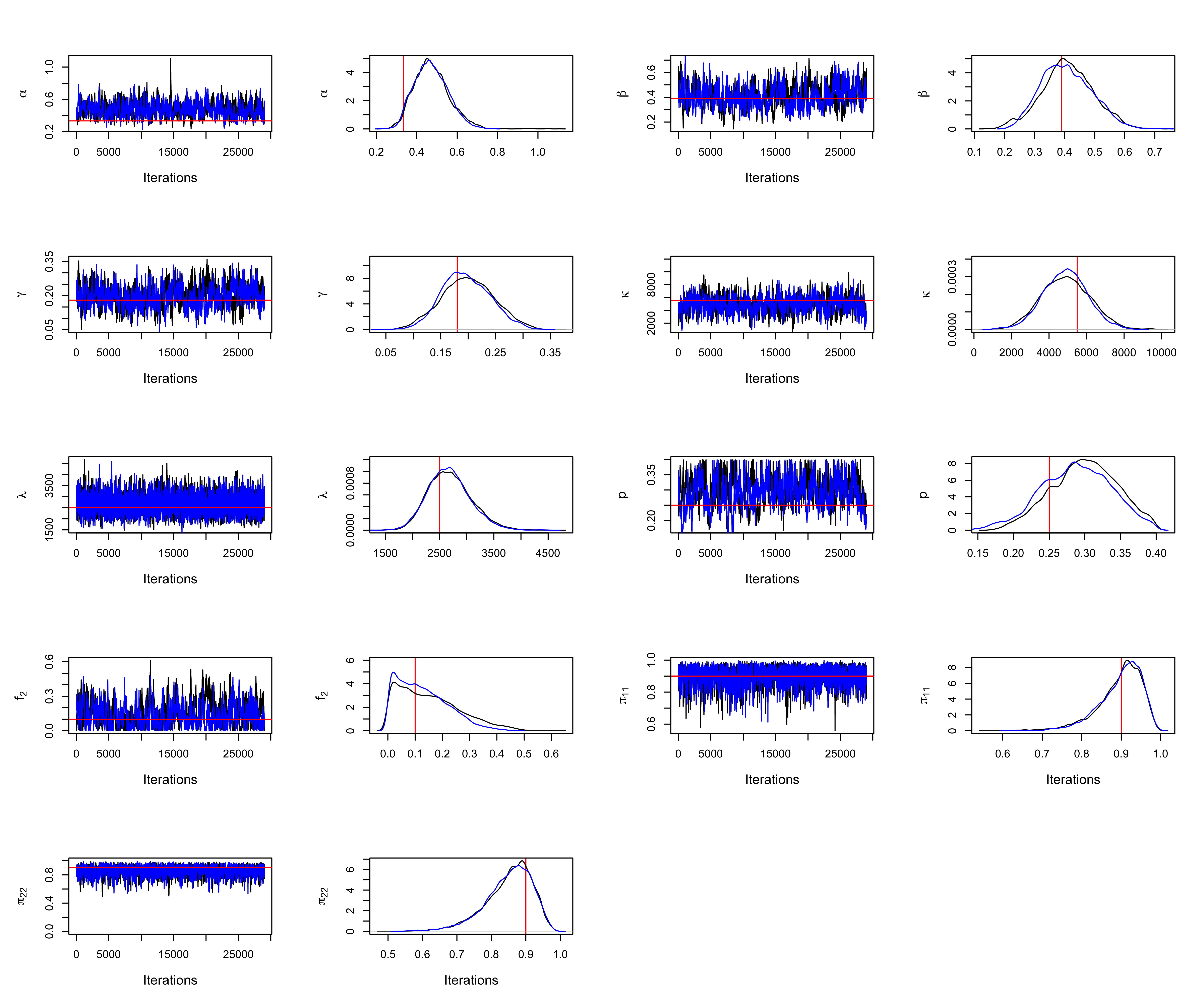}
\caption{Trace plots \added{and kernel density plots} for model parameters in the two-regime simulation study. Red lines indicate true values of parameters.}
\label{fig: two-regime simulation trace plot}
\end{figure}

\begin{figure}[ht!]
    \centering
    \includegraphics[scale=0.075]{ 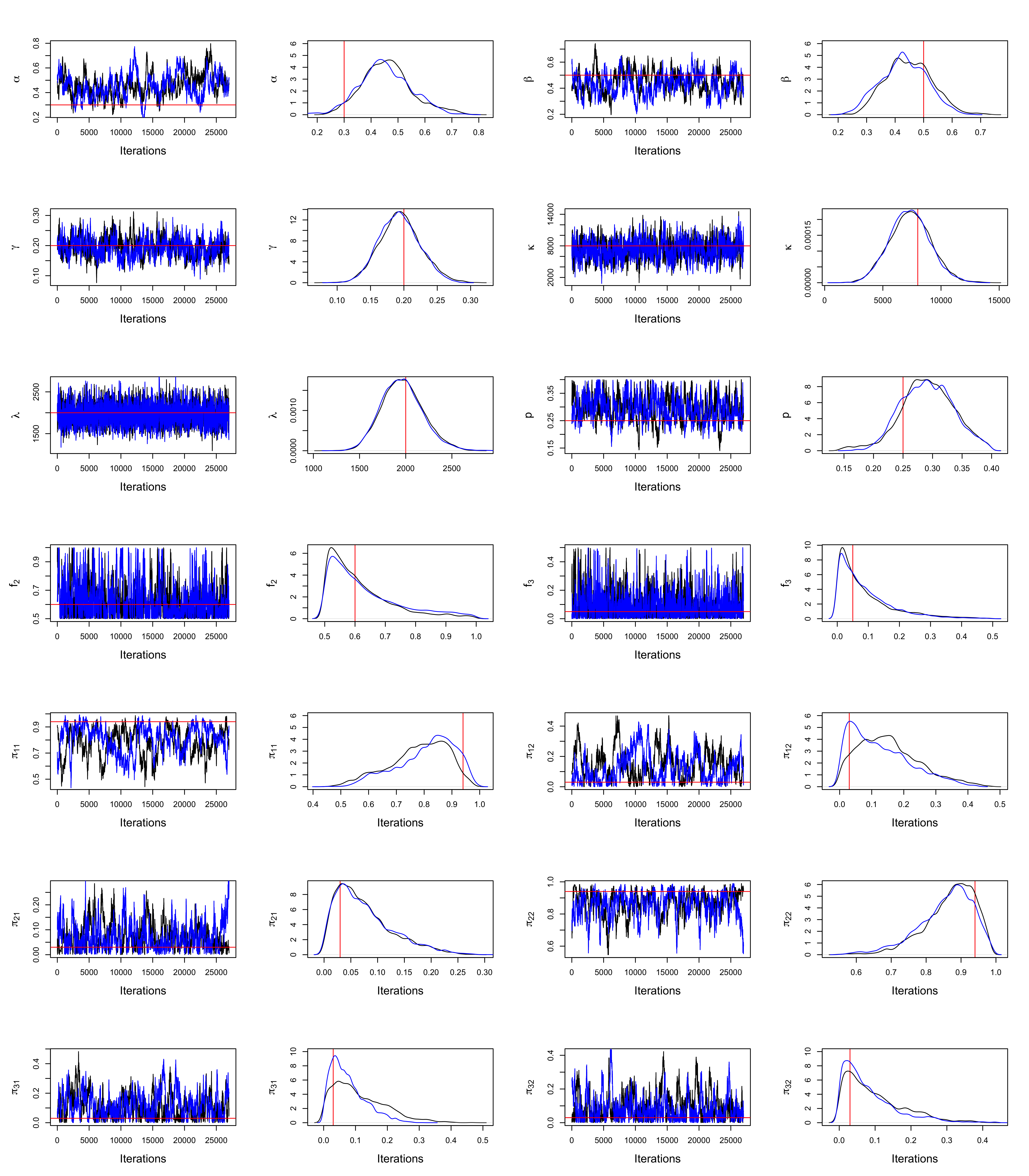}
    \caption{Trace plots \added{and kernel density plots} for model parameters in the three-regime simulation study. Red lines indicate true values of parameters. }
    \label{fig: three-regime simulation trace plot}
\end{figure}

\begin{figure}[ht!]
    \centering
    \includegraphics[scale=0.1]{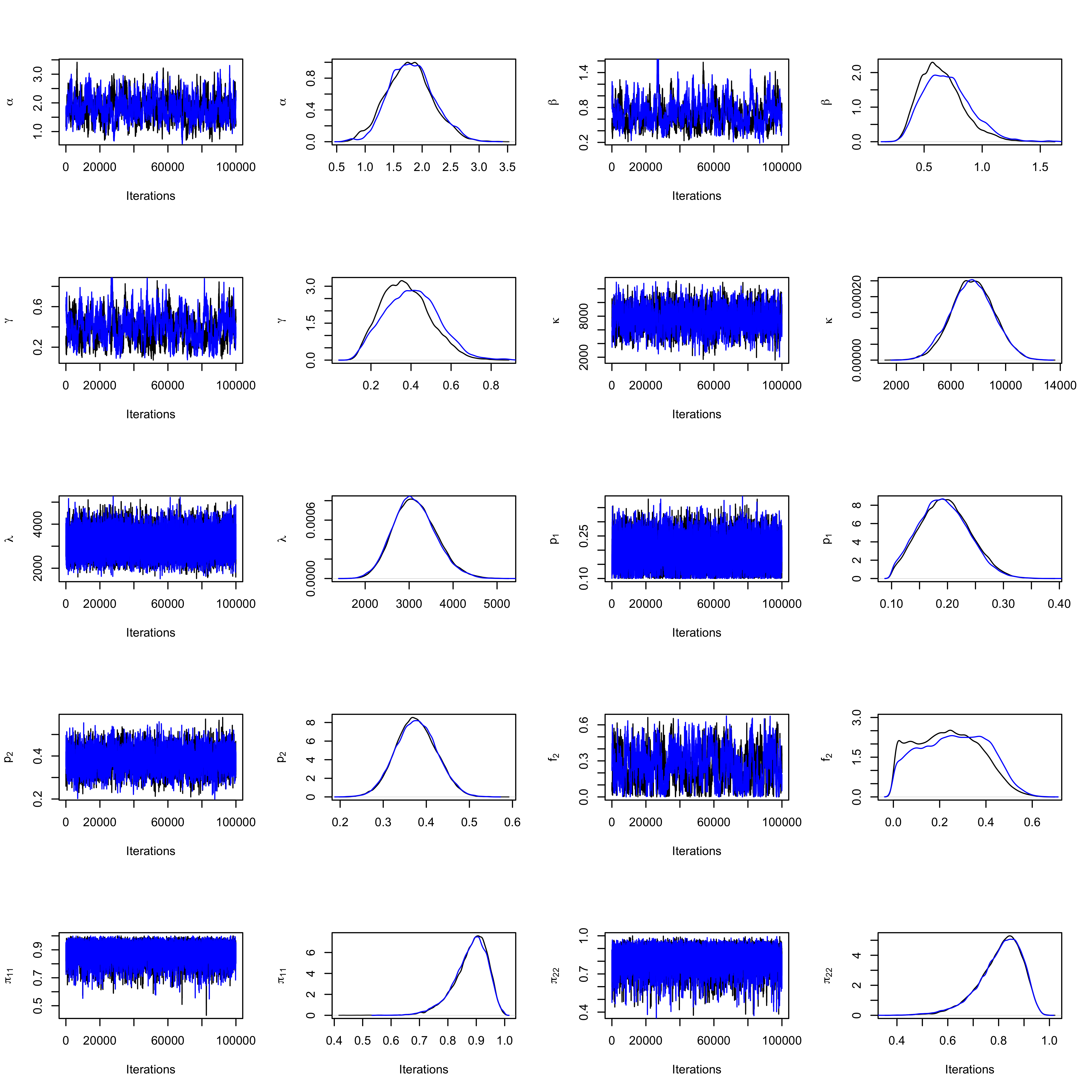}
    \caption{Trace plots \added{and kernel density plots} of model parameters based on BC COVID-19 weekly active proportions in a two-regime setting. All parameters are generated from their conditional posterior distribution by Metropolis-Hastings within particle Gibbs. Step sizes are adjusted to ensure the acceptance rate is larger than 30\,\%. }
    \label{fig: two-regime real data analysis trace plot}
\end{figure}

\clearpage
\section{Gelman-Rubin diagnostic}
\label{sec: Gelman-Rubin}
We calculate the Gelman-Rubin statistic for two Markov chains. If Gelman-Rubin statistic is less than 1.2 for all model parameters, as suggested by \cite{brooks1998general}, we can be fairly confident that convergence has been reached.

\begin{table}[ht] 
\centering
\begin{tabular}{cccccccccc}
\hline
Parameters              & $\alpha$ & $\beta$ & $\gamma$ & $\kappa$ & $\lambda$ & $p$  & $f_2$ & $\pi_{11}$ & $\pi_{22}$ \\ \hline
Gelman-Rubin Statistic & 1.00        & 1.03    & 1.05     & 1.01     & 1.00         & 1.01      & 1.13     & 1.00         & 1.00          \\ \hline
\end{tabular}
\caption{Gelman-Rubin diagnostic for the posterior distribution of parameters in the two-regime simulation study. Parameters with Gelman-Rubin close to 1 suggest good convergence. } 
\label{table: GR for two-regime simulation study}
\end{table} 

\begin{table}[ht]
\centering
\begin{tabular}{ccccccccc}
\hline
Parameters              & $\alpha$ & $\beta$ & $\gamma$ & $\kappa$ & $\lambda$ & $p$  & $f_2$ & $f_3$ \\
\hline
Gelman-Rubin Statistic & 1.04        & 1.01    & 1.01    & 1.00     & 1.00         & 1.01      & 1.00     & 1.00 \\
\hline
Parameters       & $\pi_{11}$ & $\pi_{12}$ & $\pi_{21}$ & $\pi_{22}$ & $\pi_{31}$ & $\pi_{32}$ \\
\hline
Gelman-Rubin Statistic & 1.11        & 1.05    & 1.01    & 1.06     & 1.07         & 1.14 \\
\hline
\end{tabular}
\caption{Gelman-Rubin diagnostic for the posterior distribution of parameters in the three-regime simulation study. Parameters with Gelman-Rubin close to 1 suggest good convergence.} 
\label{table: GR for three-regime simulation study}
\end{table}

\begin{table}[ht]
\centering
\begin{tabular}{ccccccccccc}
\hline
Parameters              & $\alpha$ & $\beta$ & $\gamma$ & $\kappa$ & $\lambda$ & $p_1$ & $p_2$ & $f_2$ & $\pi_{11}$ & $\pi_{22}$ \\ \hline
Gelman-Rubin Statistic & 1.01        & 1.01    & 1.01   & 1.00    & 1.00         & 1.00      & 1.00    & 1.01         & 1.00   & 1.00         \\ \hline
\end{tabular}
\caption{Gelman-Rubin diagnostic for the posterior distribution of parameters in two-regime real data analysis. Parameters with Gelman-Rubin close to 1 suggest good convergence.} 
\label{table: GR for two-regime real data analysis}
\end{table}

\end{document}